%% file: groundstate.tex
\documentclass[aps,notitlepage,onecolumn,nofootinbib,superscriptaddress,11pt,longbibliography]{revtex4-1}
\input{pretex}
\usepackage{amsmath,amsfonts,amssymb,array,graphicx,mathtools,multirow,bm,tcolorbox,relsize,booktabs}
\usepackage{tabularx}
\usepackage{subcaption} 
\usepackage{caption}
\usepackage{array}
\usepackage{xcolor}
\usepackage[utf8]{inputenc}
\usepackage[qm]{qcircuit}
\usepackage{bm}

\newcommand{\braketmatrix}[3]{\left \langle #1 \middle| #2 \middle| #3 \right \rangle}
\allowdisplaybreaks
\begin{document}
\title{Ground state preparation with shallow variational warm-start}
\author{Youle Wang}
\email{youle.wang92@gmail.com}
\author{Chenghong Zhu}
\email{zhuchenghong06@gmail.com}
\author{Mingrui Jing}
\email{mingruij0031@gmail.com}
\author{Xin Wang}
\email{wangxin73@baidu.com}

\affiliation{Institute for Quantum Computing, Baidu Research, Beijing 100193, China}

 \date{\today}
 
\begin{abstract}

Preparing the ground states of a many-body system is essential for evaluating physical quantities and determining the properties of materials. This work provides a quantum ground state preparation scheme with shallow variational warm-start to tackle the bottlenecks of current algorithms, i.e., demand for prior ground state energy information and lack of demonstration of efficient initial state preparation. Particularly, our methods would not experience the instability for small spectral gap $\Delta$ during pre-encoding the phase factors since our methods involve only $\widetilde{O}(1)$ factors while $\widetilde{O}(\Delta^{-1})$ is requested by the near-optimal methods.
We demonstrate the effectiveness of our methods via extensive numerical simulations on spin-$1/2$ Heisenberg models. We also show that the shallow warm-start procedure can process chemical molecules by conducting numerical simulations on the hydrogen chain model. Moreover, we extend research on the Hubbard model, demonstrating superior performance compared to the prevalent variational quantum algorithms.

\end{abstract}
\maketitle

\section{Introduction}
The ground state is one crucial quantum state appearing in condensed matter physics, quantum chemistry, and combinatorial optimisation. Physical reality tends to force a quantum system to its ground state at low temperature, which can exist with stability~\cite{Feynman1956}. Therefore, preparing the exact form of the ground state becomes significant for evaluating the expectations of physical quantities and determining the properties of materials~\cite{Sakurai1995}. Classical methods have raised success in predicting matter behaviours. Based on quantum mechanics, traditional methods, including, Hartree-Fock (HF) method~\cite{Sherrill2009AnIT}, full configuration interaction~\cite{knowles198975}, density functional method~\cite{parr1980density}, and Monte Carlo simulations~\cite{blankenbecler1981monte} have boosted developments of quantum chemistry by providing accurate eigenstate evaluations. However, the exponential growth of Hilbert space and multi-determinant interactions~\cite{Cade_2020,Dagotto1994,Nelson_2019_dft} demand an explosive amount of memories leading to a restricted efficiency regime and forcing the problem into an NP-complete complexity~\cite{barahona1982computational}. The fermionic statistics also raise the sign problem, which pulls down the convergence speed of approximation methods~\cite{Gilks1995,del2006sequential,Scalapino2006NumericalSO,LeBlanc2015}. 

Quantum computer, originally proposed by Feynman~\cite{Feynman1982}, is naturally embedded in the quantum Hilbert space, indicating promising and competitive near-term applications, including solving ground states of different physical models and molecules~\cite{Peruzzo2014,Wecker_2015}. The typical results from variational quantum eigensolver (VQE) \cite{Peruzzo2014} demonstrated an excellent perspective for solving the ground state energies of small molecules~\cite{Kandala_2017}. Advanced variational ansatzes have been designed and achieved better performance predicting the physical properties and electronic behaviours~\cite{Wecker_2015,Dagotto1994}, such as charge and spin densities, metal-insulator transitions, and dynamics of distinct models, which could further energise the industrial development~\cite{Sapova2022} covering new material design~\cite{ostaszewski2021reinforcement} and studies on high-temperature superconductivity~\cite{Dallaire2018,Cade_2020}. {Another category, namely, adiabatic quantum algorithms has practically solved the ground state of the Ising model, demonstrating the potential of handling the combinatorial optimisation problems~\cite{Johnson2011}.} 

{One of the most promising technical routes to prepare the ground state is to utilise phase estimation.} 
{Under a reasonable assumption that the initial state $\rho$ has a non-zero overlap $\gamma$ regarding the exact ground state and the ground state energy (within the required precision) is known, one can directly apply phase estimation \cite{https://doi.org/10.48550/arxiv.quant-ph/9511026} together with amplitude amplification \cite{Brassard_2002} can project out the ground state from $\rho$ with a probability proportional to $\gamma^2$.}
However, phase estimation usually demands an extensive ancilla system, making the realisation of preparing high-fidelity states challenging. Recently, \citet{Ge2019} proposed a method to exponentially and polynomially improve the run-time on the allowed error $\eps$ and the spectral gap $\Delta$ and the initial state overlap $\gamma$ dependencies, respectively, where $\Delta$ is the difference between the ground and the first excited state energies. The same authors also discussed that polylogarithmic scaling in $\eps^{-1}$ could be attained using the method proposed by \citet{DPoulin_2009} requiring more ancilla qubits. 

Quantum singular value transformation (QSVT) \cite{Gilyn2019} is another promising route for realising the ground state preparation. Notably, \citet{Lin2019a} have shown a method using constant number of ancilla qubits to achieve a quantum complexity $\mathcal{O}(\Delta^{-1}\gamma^{-1}\log(\eps^{-1}))$\footnote{The normalisation factor $\alpha$ in the block encoding is omitted.}. Their algorithms use the block encoding input model of the Hamiltonian \cite{Berry2015} rather than the evolution operator used in phase estimation. They also provide lower bounds of quantum complexity to show the algorithms' dependence optimality on the spectral gap and initial overlap. Subsequently, \citet{Dong2022} proposed a technique called ``quantum eigenvalue transformation of unitary matrices" (QET-U), which is closely related to quantum signal processing \cite{Low2017} and QSVT \cite{Gilyn2019}. With QET-U and controlled Hamiltonian evolution, they further reduce the ancilla qubits to two while reaching the same asymptotic scaling in $\eps$, $\Delta$, $\gamma$ as \cite{Lin2019a}. Remarkably, both \cite{Lin2019a} and \cite{Dong2022} assume an upper bound $\mu$ of ground state energy at priority. If no such $\mu$ is known, it can be obtained by first employing ground state energy estimation \cite{Abrams_1999,Higgins_2007,Berry_2009,DPoulin_2009,PhysRevA.75.012328,Lin2019a,Dong2022,Lin2022,https://doi.org/10.48550/arxiv.2211.11973,O_Brien_2019,PhysRevLett.122.140504,https://doi.org/10.48550/arxiv.1907.11748,Ge2019,Wan_2022,wang2022quantum} with precision $\mathcal{O}(\Delta)$, which would increase the cost of state preparation.

\begin{figure}[t]
    \centering
    \includegraphics[width=1\textwidth]{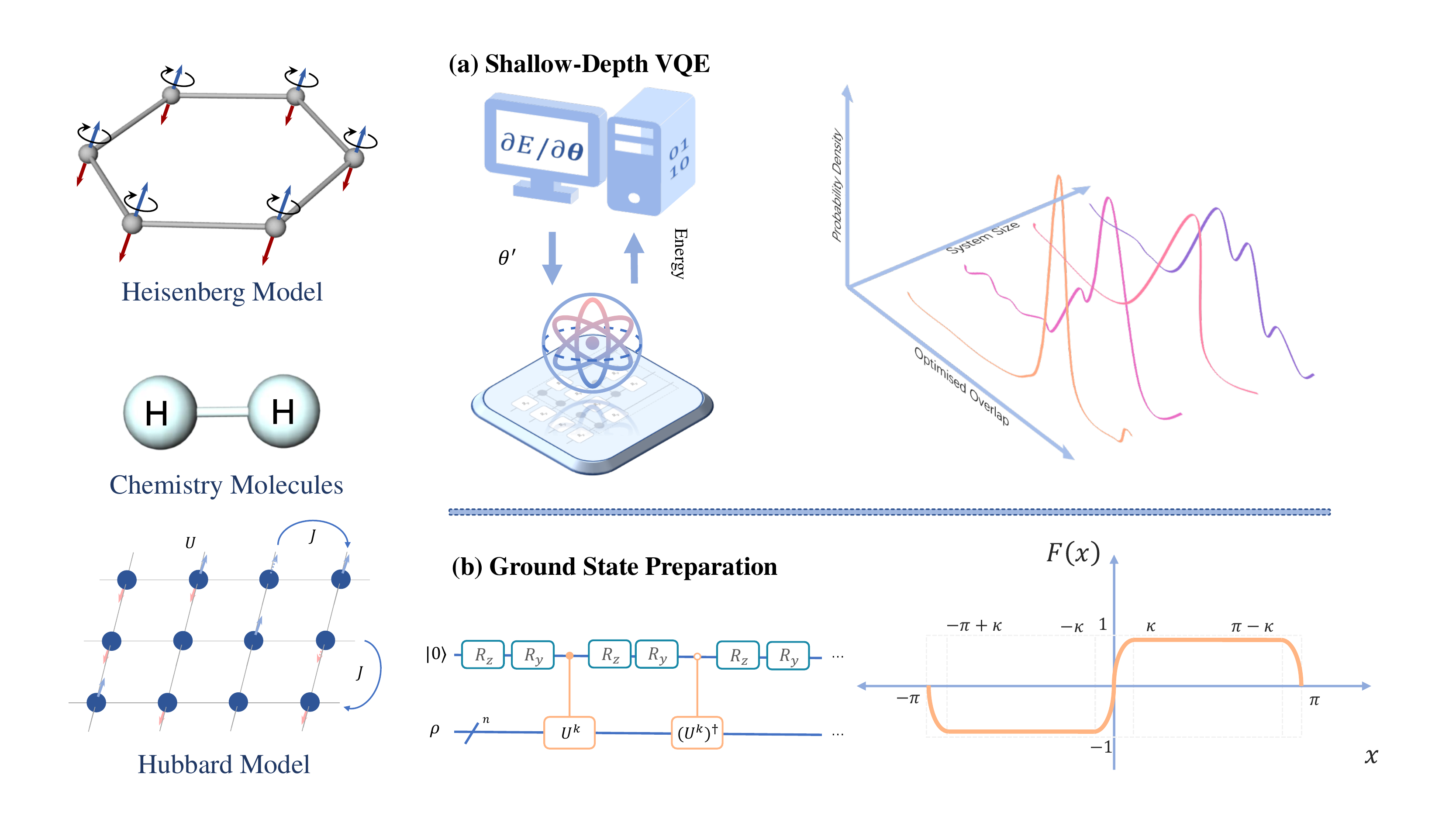}
    \caption{The outline of our ground state preparation with shallow variational warm-start involving (left) the three applicable physical models of {Heisenberg spin-$1/2$ chain model, chemical molecular model and Fermi-Hubbard model} could be efficiently solved via our scheme; (middle) the two main components of our scheme of (a) shallow-depth VQE as the warm-start and (b) the ground state preparation algorithms provided the warm-start initial state; (right) our demonstrations from testing (top) warm-start with a massive number of models and (bottom) the principle of the ground state preparation. The sign function $F(x)$ is simulated via the ground state preparation circuit.}
    \label{fig:algorithm_process}
\end{figure}

To solve ground-state problems in physics and chemistry, we notice the demand for overcoming limitations set by the current algorithms, such as the prior information of the ground-state energy and an efficient initial state preparation. This paper addresses these issues using shallow-depth VQE and the recent advances in quantum phase estimation \cite{Wang2022}. Intuitively, one would expect to prepare the ground state with very high fidelity when estimating the ground state energy. However, it cannot be realised using the conventional phase estimation due to the imperfect residual states, as discussed in \cite{DPoulin_2009,Ge2019}. Fortunately, this expectation can be fulfilled by the algorithm of \cite{Wang2022}, which uses only one ancilla qubit and can apply to block encoding and Hamiltonian evolution input models. We show that measuring the ancilla qubit can estimate the ground state energy up to the desired precision with a probability $\approx \gamma^2$. Setting the precision as $\mathcal{O}(\Delta)$, the ground state can be prepared up to exponentially high precision at a cost $\widetilde{\mathcal{O}}(\gamma^{-2}\Delta^{-1})$\footnote{Notation $\widetilde{\mathcal{O}}$ hides the logarithmic factors.}. As a notable property, the prior ground state energy information is no longer necessary. 

\begin{table}[t]
\renewcommand\arraystretch{1.7}
\centering
 \tabcolsep 3pt
\begin{tabular}{c||c|c|c|c|c|c}
\hline\hline
\multirow{2}{*}{Methods} & Input  & \#Phase  & Query & \multirow{2}{*}{\#Ancilla Qubits} & Need prior &   Need amp.   \\
& Model  & Factors & Complexity &  & GSE info.? &  amplif.? \\
\hline \hline
 \multirow{2}{*}{This work} & HE  &\multirow{2}{*}{$\widetilde{\mathcal{O}}(1)$}  & $\widetilde{\mathcal{O}}(\Delta^{-1}\gamma^{-2})$ & $1$ &  \multirow{2}{*}{No} & \multirow{2}{*}{No} \\\cline{4-4}\cline{5-5}
\cline{2-2}
  & BE &    & $\widetilde{\mathcal{O}}(\alpha\Delta^{-1}\gamma^{-2})$ & 2 & & \\ 
\hline
 \multirow{2}{*}{\citet{Dong2022}} &\multirow{2}{*}{HE}  & \multirow{2}{*}{$\widetilde{\mathcal{O}}(\Delta^{-1})$} & $\widetilde{\mathcal{O}}(\Delta^{-1}\gamma^{-2})$ &  1 & \multirow{2}{*}{Yes} & No \\
\cline{4-4} \cline{5-5} \cline{7-7}
 & &  & $\widetilde{\mathcal{O}}(\Delta^{-1}\gamma^{-1})$  & 2 & & Yes \\
\hline
\multirow{2}{*}{\citet{Lin2019a}} & \multirow{2}{*}{BE}   & \multirow{2}{*}{$\widetilde{\mathcal{O}}(\Delta^{-1})$}  &  \multirow{2}{*}{$\widetilde{\mathcal{O}}(\alpha\Delta^{-1}\gamma^{-1})$} & \multirow{2}{*}{$\mathcal{O}(1)$} &  \multirow{2}{*}{Yes} & \multirow{2}{*}{Yes} \\
 & &  &   &  & &  \\
 \hline
 \multirow{2}{*}{\citet{Ge2019}}  & \multirow{2}{*}{HE}  & \multirow{2}{*}{---}  &  \multirow{2}{*}{$\widetilde{\mathcal{O}}(\Delta^{-1}\gamma^{-1})$} & \multirow{2}{*}{$\mathcal{O}\left(\log(\Delta^{-1})+\log\log(\eps^{-1})\right)$} &  \multirow{2}{*}{Yes} & \multirow{2}{*}{Yes} \\
 & &  &   & &  &   \\
\hline\hline
\end{tabular}
\caption{Comparison to the current quantum ground-state preparation algorithms. $\Delta$ is a lower bound of the spectral gap of the Hamiltonian. Here, we use $\gamma=\sqrt{\tr(\rho\Pi)}$ to represent the overlap between the initial and ground states, where $\Pi$ denotes a projection onto the subspace generated by ground states. ``GSE" represents the ground state energy, and ``amp. amplif." is the abbreviation of amplitude amplification. ``HE" means Hamiltonian evolution $U=e^{iH}$, and ``BE" stands for block encoding $U_H=\left[\begin{smallmatrix}
    H/\alpha & \cdot \\ \cdot & \cdot
\end{smallmatrix}\right]$, where $\alpha$ is the normalisation factor. If $U_H$ is constructed by the linear-combination-of-unitaries technique \cite{childs2018toward}, the ancillas of our method can be further reduced to 1. The number of ancillas of \cite{Lin2019a} is at least three, which are used to block-encode the projector by QSVT. The ancillas for implementing amplitude amplification need to be counted as well. 
}
\label{table:ground state preparation}
\end{table}

In our algorithms, we apply a sequence of single-qubit rotations on the ancilla qubit and interleave the power of the input models. The structures of QSVT and QET-U are similar to ours but use block encoding and Hamiltonian evolution, respectively. Rotation angles, usually referred to as the phase factors, are chosen so that the circuit can simulate a specific function, such as sign function \cite{Lin2019a,Wang2022} and shifted sign function \cite{Dong2022}. The number of phase factors is crucial for preparing a high-fidelity ground state. In \cite{Lin2019a,Dong2022}, the number scales as $\mathcal{O}(\Delta^{-1}\log(\eps^{-1}))$, given the spectral gap $\Delta$ and an approximation error $\eps$. However, classically computing the factors would suffer from numerical instability, especially when $\Delta$ is small. A convex-optimisation-based method is proposed in \cite{Dong2022} to generate a near-optimal approximation. In contrast, our algorithms advantageously require $\widetilde{\mathcal{O}}(1)$ number of factors, endowing better applicability in practice than \cite{Lin2019a,Dong2022}. In particular, our numerical experiments have used circuits with fixed phase factors to prepare ground states of more than 200 random Hamiltonian instances. To better understand our algorithms, we provide a schematic diagram in Fig. \ref{fig:algorithm_process} and summarize the comparisons to \cite{Ge2019,Lin2019a,Dong2022} in Table \ref{table:ground state preparation}.

Although algorithms of \cite{Lin2019a,Dong2022} have better scaling in $\gamma$, the scheme relies on amplitude amplification incurring more ancilla qubits and deeper circuits. The quantum complexity can be compensated by preparing an initial state with an overlap larger than 0.1 since the quadratic scaling in overlap would make no big difference. To this end, we have conducted extensive zero-noise numerical simulations to apply a depth-three VQE to warm-start ground-state preparation of many-body systems and find that a shallow-depth VQE could play a good overlap initialiser. Firstly, we randomly test 10000 {two-body interactions, spin-$1/2$ Heisenberg-type models} for every case $\leq$ 10 qubits, 200 and 100 models for the cases of 12 and 14 qubits, respectively. We observe that the optimised overlaps highly cluster between 0.4 and 1, and ground states with the fidelity of almost one are prepared via our algorithms. {Secondly, we investigate the shallow-depth warm-start on molecular ground state preparation offering satisfactory performances for small bond distances.}
Thirdly, we apply our methods to solve the Hubbard model which the 2D case is recognised as a classical challenge~\cite{LeBlanc2015,Scalapino2006NumericalSO,hubbardmodel2013,Cade_2020} and quantumly demonstrated only by VQE-based methods~\cite{Cade_2020,Wecker_2015}. Our scheme derives extremely accurate charge-spin densities and the chemical potentials of a specific 1D model, reaching an absolute error around $10^{-5}$ beyond the chemical accuracy with respect to the full configuration interaction (FCI) method. 

\textbf{Organisation.} In Sec.~\ref{sec:algorithm}, we give technical details of algorithms for ground state preparation and provide a method to analyse the success probability. In Sec.~\ref{sec:heisenberg}, we show extensive numerical results that demonstrate the efficacy of a constant-depth VQE in processing Heisenberg models up to 14 qubits. We also show correctness by applying the algorithms to the optimised states of VQE. In Sec.~\ref{sec:molecule}, we numerically examine the constant-depth warm-start on a hydrogen chain model. In Sec.~\ref{sec:hubbard}, we use our algorithms to study the Fermi-Hubbard models and accurately evaluate the model's physical properties. Finally, we summarize our results and discuss potential applications in Sec.~\ref{sec:conclusion}.

\section{Ground state preparation}\label{sec:algorithm}
In this section, we present ground-state preparation algorithms. The core of our approaches has two steps: firstly, prepare an initial state with a considerable overlap, such as states generated by shallow-depth VQE, Gibbs-VQE \cite{Wang2020}, or Hartree-Fock states (which will be discussed in numerical experiments); secondly, use a quantum algorithm to project out the ground state from the initial state with high probability. The projection can be realised using the quantum algorithmic tool, namely the quantum phase search algorithm (QPS) \cite{Wang2022}, which can classify the eigenvectors of a unitary operator. Using QPS, the probability of sampling the ground state is highly approximate to the initial overlap. Therefore, by repeating these two steps sufficient times, one can obtain the desired state with arbitrarily high precision.

Motivated by the previous works \cite{Dong2022,Ge2019,Lin2019a}, we establish the algorithms for the ground state preparation with two different Hamiltonian input modes: one is the real-time Hamiltonian evolution $U=e^{iH}$, and the other uses the block encoding operator \cite{Low_2019}, which is a unitary matrix $U_H=\left[\begin{smallmatrix}
    H/\alpha & \cdot \\ \cdot & \cdot
\end{smallmatrix}\right]$ with a scaled Hamiltonian in the upper-left corner. Since the size of $U_H$ is larger than that of $H$, it would require an ancilla system to implement $U_H$ on quantum computers. A common strategy to block-encode the Hamiltonian of a many-body system is the linear-combination-of-unitaries (LCU) technique \cite{childs2018toward}. We, therefore, could regard the LCU-type block encoding as the input of our algorithms. To better describe our algorithms, we first make some assumptions: the ancilla system for implementing $U_H$ is composed of $m$ qubits, the spectrum of the matrix $H$ is restricted in the region $(-\pi, \pi)$, the spectral gap of $H$ is larger than $\Delta>0$, and $\|H\|\leq\alpha$ regarding the spectral norm. With these, the results of the algorithms are summarized in Theorem \ref{thm:main_result}.
\begin{theorem}[Ground state preparation]
    \label{thm:main_result}
    Suppose a Hamiltonian of the form $H=\sum_{j=0}^{2^n-1}\lambda_j\op{\psi_j}{\psi_j}$, where $-\pi\leq\lambda_0\leq \lambda_1\leq\ldots\leq \lambda_{2^{n}-1}\leq\pi$. Let $\Delta>0$ be a lower bound of the spectral gap. Assume access to a quantum state $\rho$ that has a non-zero overlap $\gamma=\sqrt{\tr(\rho\Pi)}$, where $\Pi$ denotes a projection operator onto the ground state subspace. Then there exists a quantum algorithm that prepares the ground state with a precision of at least $1-\eps$.
    \begin{itemize}
        \item If the Hamiltonian is accessed through a Hamiltonian evolution operator $U=e^{iH}$, then the total queries to controlled $U$ and $U^\dagger$ is $\mathcal{O}(\gamma^{-2}\Delta^{-1}\log(\eps^{-1}\log(\Delta^{-1})))$. The number of used ancilla qubits is $1$ and used copies of $\rho$ is  $\mathcal{O}(\gamma^{-2})$. 
        \item If the Hamiltonian is accessed through a block encoding operator $U_H=\left[\begin{smallmatrix}
    H/\alpha & \cdot \\ \cdot & \cdot
\end{smallmatrix}\right]$ that satisfies $\braketmatrix{0^m}{U_H}{0^m}=H/\alpha$, then the total queries to controlled $U_H$ and $U_H^\dagger$ is $\mathcal{O}(\alpha\gamma^{-2}\Delta^{-1}\log(\eps^{-1}\log(\alpha\Delta^{-1})))$. The number of used ancilla qubits is $2$ and used copies of $\rho$ is $\mathcal{O}(\gamma^{-2})$.
    \end{itemize}
\end{theorem}
The trace distance of quantum states characterises the precision, and the notation $\ket{0^m}$ really indicates $\ket{0^m}\otimes I$, where we have omitted the identity operator on the main system. 

Next, we concretely discuss the algorithm using Hamiltonian evolution in Sec. \ref{sec:he}. We provide the detailed procedure of finding the ground state via QPS and the analysis of the output probability. The algorithm using block encoding is further discussed in Sec. \ref{sec:be}. Since the analysis of the algorithm's correctness is similar from the previous literature, we mainly discuss the differences from using the Hamiltonian evolution. 

\subsection{Algorithm using real-time Hamiltonian evolution}\label{sec:he}
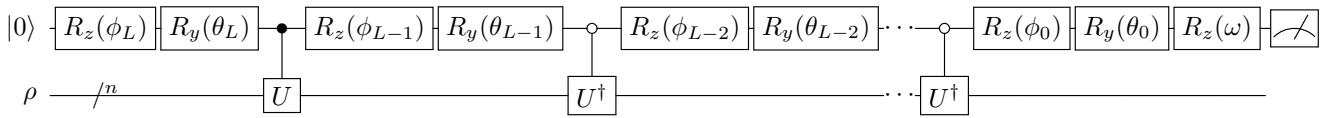
\begin{figure}[htb]
\[ 
\Qcircuit @C=0.2em @R=1em {
\lstick{\ket{0}} & \gate{R_z(\phi_L)} & \gate{R_y(\theta_L)} & \ctrl{1}  & \gate{R_z(\phi_{L-1})} & \gate{R_y(\theta_{L-1})} & \ctrlo{1} & \gate{R_z(\phi_{L-2})} & \gate{R_y(\theta_{L-2})} & \qw & &&\cdots&&& & \ctrlo{1} & \gate{R_z(\phi_{0})} & \gate{R_y(\theta_{0})} &\gate{R_z(\omega)} &\qw &\meter\\
\lstick{\rho} &/^n\qw & \qw & \gate{U}  & \qw & \qw  & \gate{U^\dagger} & \qw & \qw & \qw & &&\cdots&&& & \gate{U^\dagger} & \qw & \qw & \qw &\qw
}
\]
\caption{Quantum circuit for quantum phase search. The unitary $U$ would be adaptively changed with its power. Parameters $\bm\phi$, $\bm\theta$, and $\omega$ are chosen such that the circuit can implement a trigonometric polynomial transformation to each phase of the unitary operator. The integer $L$ is the order of the polynomial.
}
\label{fig:circuit of QPP}
\end{figure}

Given access to the controlled versions of the unitary $U=e^{iH}$ and $U^\dagger=e^{-iH}$, we construct the quantum circuit of QPS as shown in Fig.~\ref{fig:circuit of QPP}. 
There is an ancilla system consisting of only one qubit, which is applied with a sequence of single-qubit rotations. Meanwhile, several controlled unitaries are interleaved between rotations. The phase factors $\bm\phi$, $\bm\theta$, and $\omega$ are chosen such that the circuit implements a trigonometric polynomial transformation to each phase of the unitary, i.e., $f: \lambda \mapsto \sum_{j=-L}^{L}c_je^{ij\lambda}$, where each $c_j\in\mathbb{C}$ and satisfies $\sum_{j}|c_j|\leq 1$. For a certain trigonometric polynomial, the corresponding factors can be classically computed beforehand \cite{Dong_2021,Wang2022}. The action of running the circuit to an eigenvector of the unitary $U$ is presented in the lemma below.
 
\begin{lemma}[Quantum phase evolution \cite{Wang2022}]\label{lem:phase classification}
Given a unitary operator $U = \sum_{j=0}^{2^n - 1} e^{i \lambda_j} \ketbra{\psi_j}{\psi_j}$, for any trigonometric polynomial $f(x)=\sum_{j=-L}^{L}c_je^{ijx}$ with $\sum_{j}|c_j|\leq 1$, there exist parameters $\omega \in \RR$ and $\bm\theta,\bm\phi \in \RR^{L+1}$ such that the circuit in Fig. \ref{fig:circuit of QPP} transforms the initial state $\ket{0,\psi}$ into $\left(\sqrt{(1+f(\lambda))/2} \ket{0} + \sqrt{(1-f(\lambda))/2} \ket{1}\right)\ket{\psi}$, 
where $\ket{\psi}$ denotes an eigenvector of $U$ with a phase $\lambda\in(-\pi,\pi)$.
\end{lemma}

For QPS, the trigonometric polynomial $f(x)=\sum_{j=-L}^{L}c_je^{ijx}$ is chosen to approximate the sign function. Technically, supposing a constant $\kappa\in(0,1/2)$, we could find a $f(x)$ with order $L\in\mathcal{O}(\kappa^{-1}\log(\eps^{-1}))$ such that $|f(x)-1|\leq \eps$ for all $x\in  (\kappa, \pi-\kappa)$ and $|f(x)+1|\leq \eps$ for all $x\in (-\pi + \kappa, -\kappa)$ \cite{https://doi.org/10.48550/arxiv.1707.05391,Wang2022}. Particularly, given the relation between the order $L$ and precision $\eps$, we could choose $\eps$ to be exponentially small at the expense of a modest increase in the circuit depth. Therefore, eigenvectors fall in the region $(-\pi + \kappa, -\kappa)\cup (\kappa, \pi-\kappa)$ are labelled by states 0 or 1, and the collapsed state from the measured ancilla qubit could indicate the label. For instance, for any state $\ket{\psi}=\sum_{j}\alpha_j\ket{\psi_j}$, the circuit of QPS transforms the initial state $\ket{0,\psi}$ into the following form.
\begin{align}
\ket{0}\sum_{j}\alpha_j\sqrt{\frac{1+f(\lambda_j)}{2}}\ket{\psi_j}+\ket{1}\sum_{j}\alpha_j\sqrt{\frac{1-f(\lambda_j)}{2}}\ket{\psi_j}.
\end{align}
We readily know that $\ket{0}$ is associated with eigenvectors in the region $(-\kappa, \pi+\kappa)$, and $\ket{1}$ is associated with eigenvectors in the region $(-\pi-\kappa, \kappa)$. The only possible source of error comes from the intrinsic sign function approximating error from the searching procedures discussed previously, which raises negligible effect. This property enables one to efficiently locate the region where the ground state energy falls through a binary search.

\subsubsection*{Rough ground state energy search}\label{subsubsec:ground state search}
By executing the binary search, the region containing the ground state energy would shrink gradually, and the probability of outputting the ground state is analyzed later. After each iteration, the underlying region is cut into two halves, and one of them is chosen, where the post-measurement state falls in, for the next iteration. Since the region's size decreases by nearly half, the search could halt in a logarithmic time, producing an exponentially-small region. When the final region of the ground state is attained, any point in the final region could be taken as an estimate of ground state energy. In particular, if the estimation accuracy is smaller than half of the spectral gap, the ground state would eventually fall into a region whose size is smaller than the spectral gap, and excited states would be filtered out. At that time, the post-measurement state on the primary system is the desired ground state. 

To better understand the algorithm, we sketch the search procedure below. For remark, the final post-measurement state is not the desired ground state. We defer the discussion of improving the estimation accuracy.
\begin{enumerate}
    \item Input: unitary $U=e^{iH}$, constant $\kappa$ (irrelevant to $H$), error tolerance $\eps$, and state $\rho$ that non-trivially overlaps with the ground state.
    \item Compute parameters $\bm\phi$, $\bm\theta$, and $\omega$ according to parameters $\kappa$ and $\eps$.
    \item Choose the initial region $(x_{0}, x_{1})$, where $x_{0}=-\pi$ and $x_{1}=\pi$.
    \item Calculate the middle point of the region $(x_{0}, x_{1})$, which is $x_{\rm mid}=\frac{x_{0}+ x_{1}}{2}$.
    \item Set the circuit of QPS using controlled $\tilde{U}=e^{-ix_{\rm mid}}U$ and $\tilde{U}^\dagger$ and the calculated parameters from the step 2.
    \item Input the initial state $\op{0}{0}\otimes \rho$ and run the circuit.
    \item Measure the ancilla qubit of the circuit and update the region according to the measurement outcome: when $x_1-x_0>2\pi-2\kappa$, execute the update
\begin{equation}
\left\{
\begin{matrix}
(x_0,x_1)\leftarrow (x_{\rm mid}-\kappa, x_1 +\kappa), & \textrm{if the outcome is 0;}  \\[0.5em]
(x_0,x_1)\leftarrow (x_0-\kappa, x_{\rm mid}+\kappa), & \textrm{if the outcome is 1.}
\end{matrix}    
\right.
\end{equation}
When $x_1-x_0\leq 2\pi-2\kappa$, execute the update
\begin{equation}
\left\{
\begin{matrix}
(x_0,x_1)\leftarrow (x_{\rm mid}-\kappa, x_1), & \textrm{if the outcome is 0;}  \\[0.5em]
(x_0,x_1)\leftarrow (x_0,x_{\rm mid}+\kappa), & \textrm{if the outcome is 1.}
\end{matrix}    
\right.
\end{equation}
    \item Go to step 4 and repeat till the region converges.
    \item Output: final region $(x_0, x_1)$.
\end{enumerate}

After executing the rough search procedure, the size of the region would converge to $2\kappa$. Specifically, after the first iteration, the size of the region is $\kappa+\pi/2$. Since then, the region is updated as $(\theta-\kappa, x_1)$ or $(x_0, \theta+\kappa)$, and the size of the new region is $\kappa+(x_1-x_0)/2$. Then we can inductively find the size of the final region. Let $(x_0^{(t)}, x_{1}^{(t)})$ denote the region in the $t$-iteration and $Q$ denote the total number of iterations, then we have
\begin{align}
|x_{1}^{(Q)}-x_{0}^{(Q)}|&=\kappa+\frac{|x_{1}^{(Q-1)}-x_{0}^{(Q-1)}|}{2}=\kappa+\frac{1}{2}\left[\kappa+\frac{|x_{1}^{(Q-2)}-x_{0}^{(Q-2)}|}{2}\right]=\cdots=2\kappa+\frac{\pi}{2^Q}.
\end{align}

\subsubsection*{Improving search accuracy} 
Suppose the ground state energy falls in the region $(x_0^{(Q)}, x_{1}^{(Q)})$, then the above procedure would return a rough estimate of the ground state energy up to precision $\kappa+\pi/2^{Q+1}$. However, the estimation precision is not guaranteed to be smaller than $\Delta/2$. Thus more processes are needed to improve the precision. It is realised by applying the power of the unitary operator that amplifies the phases of the post-measurement state. By repeatedly running the circuit with the unitary power, the ground state energy is efficiently constrained to a smaller region. To be specific, firstly, run the rough search procedure once with unitary $U^{(0)}=U$, attaining a region of size $2\kappa+\pi/2^Q$. For clarity, denote this region by $(\zeta_0^{(1)}, \zeta_1^{(1)})$ and $\bar{\kappa}=\kappa+\pi/2^{Q+1}$, where $|\zeta_1^{(1)}-\zeta_0^{(1)}|= 2\bar{\kappa}$. Let $\lambda^{(1)}=(\zeta_0^{(1)}+\zeta_1^{(1)})/2$ denote the middle point of the region. Secondly, consider a modified unitary $U^{(1)}=(e^{-i\lambda^{(1)}}U^{(0)})^{\lfloor1/\bar{\kappa}\rfloor}$\footnote{The notation $\lfloor x\rfloor$ denotes the floor function that gives as output the greatest integer less than or equal to $x$.}. For $U^{(1)}$, phases of eigenvectors in the post-measurement state are rescaled to $(-1, 1)$. Now, running the rough search procedure with the new unitary operator would give a new region $(\zeta_0^{(2)}, \zeta_1^{(2)})$. 

Executing the above two steps would exponentially fast improve the estimation accuracy {compared with the rough searching scheme.} 
For state $\ket{\psi_0}$, let $\lambda_0$ denote its original phase in $U$. Its phase in unitary $U^{(1)}$ is $(\lambda_0-\lambda^{(1)})\lfloor1/\bar{\kappa}\rfloor$ and falls in the region $(\zeta_0^{(2)},\zeta_1^{(2)})$. Let $\lambda^{(2)}=(\zeta_0^{(2)}+\zeta_1^{(2)})/2$ denote the middle point of the second region $(\zeta_0^{(2)}, \zeta_1^{(2)})$. Then we readily give an inequality that characterises the estimation error $ \left|\lambda^{(2)}-(\lambda_0-\lambda^{(1)})\lfloor1/\bar{\kappa}\rfloor\right|\leq \bar{\kappa}$. Rewrite this inequality as $\left|\lambda_0-\left(\lambda^{(1)}+\lambda^{(2)}/\lfloor1/\bar{\kappa}\rfloor\right)\right|\leq \bar{\kappa}/\lfloor1/\bar{\kappa}\rfloor\leq \lfloor1/\bar{\kappa}\rfloor^{-2}$. From this equation, we can see that the first two steps of the scheme give an estimate of the phase with an error of $\lfloor1/\bar{\kappa}\rfloor^{-2}$. So, inductively repeating the above procedure leads to a sequence $(\lambda^{(1)}, \lambda^{(2)},\ldots)$ that could be taken as an estimate of the ground state energy. Assuming repeats $j$ times, the estimate is given by
\begin{align}
    \left|\lambda_0-\left(\lambda^{(1)}+\frac{\lambda^{(2)}}{\lfloor1/\bar{\kappa}\rfloor}+\frac{\lambda^{(3)}}{\lfloor1/\bar{\kappa}\rfloor^2}+\cdots+\frac{\lambda^{(j)}}{\lfloor1/\bar{\kappa}\rfloor^{j-1}}\right)\right|\leq \frac{\bar{\kappa}}{\lfloor1/\bar{\kappa}\rfloor^{j-1}}\leq \lfloor1/\bar{\kappa}\rfloor^{-j}.
\end{align}
Clearly, reaching the halt condition only requires repeating at most $\mathcal{O}(\log(1/\Delta)/\log(\lfloor1/\bar{\kappa}\rfloor))$ times.

\subsubsection*{Output probability analysis} 
Due to many intermediate measurements in QPS, the search procedure is conducted randomly. The event that the ground state is output would occur with a probability. However, how the analysis of this probability is missing in \cite{Wang2022}. We use an example to explain how to analyse the output probability, which can be used in more general cases. 

\begin{figure}[htb]
\centering
\begin{tikzpicture}
\node [circle,draw]{$\rho$} [level distance=10mm,sibling distance=25mm]
child { node [circle,draw]{$\rho_{1,1}$} [level distance=10mm ,sibling distance=15mm]
child {node [circle,draw] {$\rho_{2,1}$}
child {node [circle,draw] {$\rho_{3,1}$}}
}
child {node [circle,draw]{$\rho_{2,2}$} [level distance=10mm ,sibling distance=10mm]
child {node [circle,draw]{$\rho_{3,2}$}}
}}
child {node [circle,draw] {$\rho_{1,2}$} [level distance=10mm ,sibling distance=10mm]
child {node [circle,draw]{$\rho_{2,3}$}
child {node [circle,draw] {$\rho_{3,3}$}}
child {node [circle,draw] {$\rho_{3,4}$}}
}
};
\end{tikzpicture}
\caption{An example of a binary tree diagram for sketching phase search. The node with notation $\rho_{l,k}$ represents the $k$th post-measurement state at $l$ iteration. }
\label{fig:QPS_binary_tree}
\end{figure}
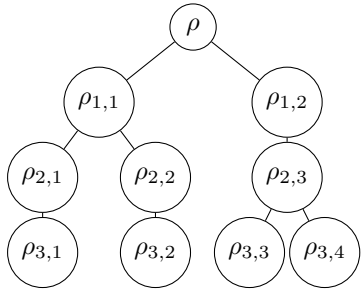

\textbf{Example.} 
If we collect all measurement outcomes together to represent a trajectory that the ground state moves along, the search process of QPS can be depicted using a binary tree as shown in Fig.~\ref{fig:QPS_binary_tree}. The root node represents the initial state $\rho$, and other nodes represent the post-measurement state with a non-zero overlap concerning the ground state. Each leaf node in the tree {represents a possible searched ground state from the program}. Each trajectory consists of four nodes, meaning that QPS finds the ground state after three iterations. In addition, the line goes left, meaning that the measurement outcome is 0, and goes right, meaning that the measurement outcome is 1. 

By the law of total probability, the probability of finding the ground state is the sum of the probability of following each trajectory. We take the most left trajectory that reaches the leaf node $\rho_{3,1}$ as an example to explain how to compute the probability of a trajectory. Let $\Pr[\rho_{i,j}|\rho_{p,l}]$ denote the conditional probability of reaching node $\rho_{i,j}$ from node $\rho_{p,l}$. Clearly, the probability that $\rho_{p,l}$ transits to $\rho_{i,j}$ only depends on the state $\rho_{p,l}$. Then the probability of reaching the leaf node $\rho_{3,1}$ from the root node can be given by 
\begin{align}
    \Pr[\rho_{3,1}]=\Pr[\rho_{3,1}|\rho_{2,1}]\times \Pr[\rho_{2,1}|\rho_{1,1}] \times \Pr[\rho_{1,1}|\rho].
\end{align}

To represent probability $\Pr[\rho_{1,1}|\rho]$, we define linear maps $\mathcal{M}_{(1,1)}^{(0)}$ that transform the input density matrix $\rho$ to an un-normalised post-measurement state, where $k\in\{0,1\}$ indicates the measurement outcome. Recalling Theorem 6 of \cite{Wang2022}, the measurement probabilities are $\Pr[0]=\frac{1}{2}(1+\sum_{j}\braketmatrix{\psi_j}{\rho}{\psi_j}f(\lambda_j))$ and $\Pr[1]=\frac{1}{2}(1-\sum_{j}\braketmatrix{\psi_j}{\rho}{\psi_j}f(\lambda_j))$. Then the action of $\mathcal{M}_{(1,1)}^{(k)}$ is described as follows.
\begin{align}
    &\mathcal{M}_{(1,1)}^{(0)}:\rho=\sum_{l}p_{l}\op{\phi_l}{\phi_l} \mapsto \sum_{l}p_{l}\op{\widetilde{\phi}_l^{(0)}}{\widetilde{\phi}_l^{(0)}}, \quad \mathrm{where} \quad \ket{\widetilde{\phi}_l^{(0)}}=\sum_{j}\sqrt{\frac{1+f(\lambda_j)}{2}}\braket{\psi_j}{\phi_l}\ket{\psi_j}. \label{eq:m0}\\
    &\mathcal{M}_{(1,1)}^{(1)}:\rho=\sum_{l}p_{l}\op{\phi_l}{\phi_l} \mapsto \sum_{l}p_{l}\op{\widetilde{\phi}_l^{(1)}}{\widetilde{\phi}_l^{(1)}}, \quad \mathrm{where} \quad \ket{\widetilde{\phi}_l^{(1)}}=\sum_{j}\sqrt{\frac{1-f(\lambda_j)}{2}}\braket{\psi_j}{\phi_l}\ket{\psi_j}.
\end{align}
Note that $\ket{\widetilde{\phi}_l^{(0)}}$ and $\ket{\widetilde{\phi}_l^{(1)}}$ are not quantum states but represent column vectors. For other probabilities, the maps are defined similarly. That is, the map that transforms $\rho_{i,j}$ to $\rho_{p,l}$ is denoted by $\mathcal{M}_{(i,j)\to(p,l)}^{(0)}$. With these notations, the probabilities are re-expressed below.
\begin{align}
    \Pr[\rho_{1,1}|\rho]&=\tr\left(\mathcal{M}_{(1,1)}^{(0)}(\rho)\right) \\
    \Pr[\rho_{2,1}|\rho_{1,1}] &= \tr\left(\mathcal{M}_{(1,1)\to(2,1)}^{(0)}(\rho_{1,1})\right)= \frac{\tr\left(\mathcal{M}_{(1,1)\to(2,1)}^{(0)}\circ\mathcal{M}_{(1,1)}^{(0)}(\rho)\right)}{\tr\left(\mathcal{M}_{(1,1)}^{(0)}(\rho)\right)} \\
    \Pr[\rho_{3,1}|\rho_{2,1}]&= \tr\left(\mathcal{M}_{(2,1)\to(3,1)}^{(0)}(\rho_{2,1})\right)=\frac{\tr\left(\mathcal{M}_{(2,1)\to(3,1)}^{(0)}\circ\mathcal{M}_{(1,1)\to(2,1)}^{(0)}\circ\mathcal{M}_{(1,1)}^{(0)}(\rho)\right)}{\tr\left(\mathcal{M}_{(1,1)\to(2,1)}^{(0)}\circ\mathcal{M}_{(1,1)}^{(0)}(\rho)\right)}
\end{align}
Eventually, the probability $\Pr[\rho_{3,1}]$ can be written as 
\begin{align}
    \Pr[\rho_{3,1}]=\tr\left(\mathcal{M}_{(2,1)\to(3,1)}^{(0)}\circ\mathcal{M}_{(1,1)\to(2,1)}^{(0)}\circ\mathcal{M}_{(1,1)}^{(0)}(\rho)\right).
\end{align}

As shown in Eq.~\eqref{eq:m0}, the circuit of QPS allocates weight to each eigenvector conditional on the state of the ancilla qubit. Specifically, each eigenvector $\ket{\psi_j}$ is allocated with a weight $\sqrt{(1+f(\lambda_j))/2}$ if the ancilla qubit is $\ket{0}$; otherwise, the weight is $\sqrt{(1-f(\lambda_j))/2}$. Since $f(x)$ is an approximation of the sign function in the sense that $|f(x)-\mathrm{sign}(x)|\leq \eps$ for all $x\in(-\pi+\kappa, -\kappa)\cup(\kappa, \pi-\kappa)$, and the approximation error $\eps$ could be exponentially small, we can see that some eigenvectors are filtered from the un-normalised post-measurement state.
\begin{equation}
    \begin{array}{cc}
     \ket{\widetilde{\phi}_l^{(0)}}=\sum_{j:\lambda_j\not\in(-\pi+\kappa, -\kappa)}\sqrt{\frac{1+f(\lambda_j)}{2}}\braket{\psi_j}{\phi_l}\ket{\psi_j},    &  \ket{\widetilde{\phi}_l^{(1)}}=\sum_{j:\lambda_j\not\in(\kappa, \pi-\kappa)}\sqrt{\frac{1-f(\lambda_j)}{2}}\braket{\psi_j}{\phi_l}\ket{\psi_j}. 
    \end{array}
\end{equation}
Hence, after the operation of $\mathcal{M}_{(2,1)\to(3,1)}^{(0)}\circ\mathcal{M}_{(1,1)\to(2,1)}^{(0)}\circ\mathcal{M}_{(1,1)}^{(0)}$, the initial state $\rho$ is filtered, and only the ground state remains. Using the linearity of the map $\mathcal{M}$ and trace, we can see that 
\begin{align}
    \Pr[\rho_{3,1}]  & = \sum_{ljk}p_l \braket{\psi_j}{\phi_l}\braket{\phi_l}{\psi_k} \tr\left(\mathcal{M}_{(2,1)\to(3,1)}^{(0)}\circ\mathcal{M}_{(1,1)\to(2,1)}^{(0)}\circ\mathcal{M}_{(1,1)}^{(0)}(\op{\psi_j}{\psi_k})\right) \\
    &= \sum_{l,j: \Pi\ket{\psi_j}\neq 0, \atop k: \Pi\ket{\psi_k}\neq 0}p_l \braket{\psi_j}{\phi_l}\braket{\phi_l}{\psi_k} \tr\left(\mathcal{M}_{(2,1)\to(3,1)}^{(0)}\circ\mathcal{M}_{(1,1)\to(2,1)}^{(0)}\circ\mathcal{M}_{(1,1)}^{(0)}(\op{\psi_j}{\psi_k})\right) \\
    &=\tr\left(\mathcal{M}_{(2,1)\to(3,1)}^{(0)}\circ\mathcal{M}_{(1,1)\to(2,1)}^{(0)}\circ\mathcal{M}_{(1,1)}^{(0)}(\Pi \rho \Pi)\right).
\end{align}
where $\Pi$ denotes the projection operator onto the subspace of the ground state. The second equality holds since excited eigenvectors would be assigned zero weight. Conducting the same process would give expressions for the rest probabilities $\Pr[\rho_{3,2}]$, $\Pr[\rho_{3,3}]$, and $\Pr[\rho_{3,4}]$. Finally, simple calculations would immediately give the probability of finding the ground state. 
\begin{align}
   \Pr[\rho_{3,1}]+\Pr[\rho_{3,2}]+\Pr[\rho_{3,3}]+\Pr[\rho_{3,4}]=\tr(\rho \Pi).
\end{align}

To consider the error of approximating the sign function, the above discussion can be reduced to the case where the initial state is an eigenvector. Then the output probability is highly approximate to $\tr(\rho \Pi)$. We refer the interested readers to \cite{Wang2022} for more complexity analysis.

{For a general case, we can sketch the search process as a binary tree and calculate the probability of finding the ground state in the same manner. Hence, we conclude that the probability is highly approximate to the initial state overlap under the assumption that the Hamiltonian evolution is simulated very accurately.}

\subsection{Algorithm using block encoding}\label{sec:be}
\begin{figure}[htb]
\[ 
\Qcircuit @C=0.2em @R=1em {
\lstick{\ket{0}} & \gate{R_z(\phi_L)} & \gate{R_y(\theta_L)} & \ctrl{1}  & \gate{R_z(\phi_{L-1})} & \gate{R_y(\theta_{L-1})} & \ctrlo{1} & \gate{R_z(\phi_{L-2})} & \gate{R_y(\theta_{L-2})} & \qw & &&\cdots&&& & \ctrlo{1} & \gate{R_z(\phi_{0})} & \gate{R_y(\theta_{0})} &\gate{R_z(\omega)} &\qw &\meter\\
\lstick{\ket{0^m}} &/^m\qw & \qw & \multigate{1}{U_{H}}  & \qw & \qw  & \multigate{1}{U_{H}^\dagger} & \qw & \qw & \qw & &&\cdots&&& & \multigate{1}{U_{H}^\dagger} & \qw & \qw & \qw &\qw \\
\lstick{\rho} &/^n\qw & \qw & \ghost{U_{H}}  & \qw & \qw  & \ghost{U_{H}^\dagger} & \qw & \qw & \qw & &&\cdots&&& & \ghost{U_{H}^\dagger} & \qw & \qw & \qw &\qw
}
\]
\caption{Block-encoding based QPS circuit. $U_H$ is a block encoding of the Hamiltonian. The second line represents the block-encoding's ancilla system of size $m$. The third line stands for the main system with $n$ qubits.}
\label{fig:QPS using block encoding}
\end{figure}
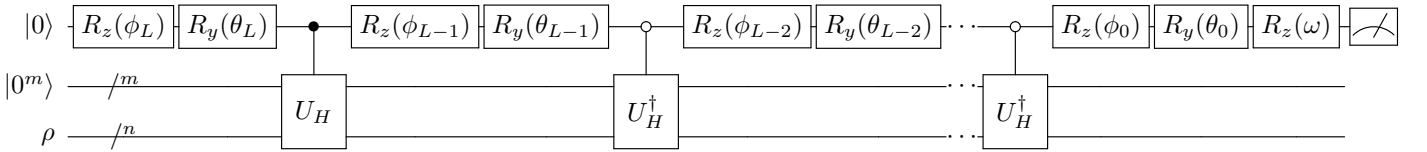

When the input mode is a block encoding operator $U_H=\left[\begin{smallmatrix}
    H/\alpha & \cdot \\ \cdot & \cdot
\end{smallmatrix}\right]$ of the Hamiltonian, QPS is similarly employed to prepare the target ground state. The discussion on probability and correctness is similar to the previous section. In contrast, there are several differences in using the Hamiltonian evolution. Firstly, implementing $U_H$ requires an ancilla system; thus, making the circuit wider. Secondly, QPS outputs the eigenvalue and eigenvector of the block encoding rather than the Hamiltonian; hence further processing is demanded. Specifically, when the QPS finishes, computing the cosine function of the output gives an estimation of the eigenvalue of the Hamiltonian. Measure the ancilla system of the final output state to achieve the ground state once the post-measurement state has the outcomes all zeros; otherwise, one has to restart the QPS procedure with the initial state or the post-measurement state and halt until all-zero outcomes are observed. Thirdly, the estimation accuracy is set smaller than $\Delta/2\alpha$. The corresponding circuit is shown in Fig. \ref{fig:QPS using block encoding}. More discussions on implementing block encoding in QPS can be found in \cite{Wang2022}.

A general block encoding cannot be directly employed to prepare the ground state unless it satisfies an invariant property. When the invariant property is satisfied, eigenvalues of $U_H$ and $H$ can be connected via a cosine function. Specifically, the action of the block encoding can be described as $U_H\ket{0^m,\psi_j}=\lambda_j/\alpha\ket{0^m,\psi_j}+\sqrt{1-(\lambda_j/\alpha)^2}\ket{\bot_j}$, where $\ket{\bot_j}$ denotes a normalised column vector satisfying $(\op{0^m}{0^m}\otimes I)\ket{\bot_j}=0$. Let $\mathcal{H}_j$ denote a two-dimensional plane spanned by vectors $\{\ket{0^m,\psi_j}, \ket{\bot_j}\}$. If $U_{H}$ is invariant in $\mathcal{H}_j$, i.e., $U_{H}: \mathcal{H}_{j} \to \mathcal{H}_j$ is a rotation, we could easily find the relation between $\lambda_j$ and the corresponding phase $\pm\tau_j$ of $U_H$, which is $\lambda_j/\alpha=\cos(\pm\tau_j)$. If $U_H$ does not satisfy the invariant property, this issue can be addressed by using the ``qubitization" technique \cite{Low_2019}. The core of qubitization is constructing a new block encoding operator using one more ancilla qubit and querying the controlled-$U_H$ and its inverse operator once. Furthermore, the LCU-type block encoding satisfies the aforementioned invariant property; hence, qubitization is no longer needed in this case.

To output an eigenvector of $U_H$ corresponding to the ground state, the estimation accuracy is set smaller than $\Delta/2\alpha$. Note that the cosine function is monotonically increasing and decreasing in the region $(-\pi,0)$ and $(0,\pi)$, respectively. We can know that the phase gap of the unitary is larger than the scaled spectral gap $\Delta/\alpha$. Thus setting accuracy as $\Delta/2\alpha$ suffices to isolate the ground state. Suppose we have attained an eigenvector, e.g., $(\ket{0^{m}, \psi_0}\pm i\ket{\bot_0})/\sqrt{2}$. Directly measuring on the ancilla system of the block encoding, the post-measurement state is the ground state if the outcomes are all zeros. The probability of this happening is exactly $1/2$. If all-zero outcomes are not observed, one can restart the whole procedure. Or one can send the post-measurement state into QPS again. Afterwards, the state will collapse to either $(\ket{0^{m}, \psi_0}+ i\ket{\bot_0})/\sqrt{2}$ or $(\ket{0^{m}, \psi_0}- i\ket{\bot_0})/\sqrt{2}$. This process is repeated until the outcomes are all zeros. Particularly, the failure probability will fast decay as $(1/2)^{r}$, where $r$ means the number of repeats.


\section{Application to Heisenberg Hamiltonians} \label{sec:heisenberg}

The general Heisenberg-typed model is arguably one of the most commonly used models in the research of quantum magnetism and quantum many-body physics that researching in the model's ground state indicates the characteristics of spin liquid and inspires crystal topology design~\cite{Yan_2011}. The Hamiltonian can be expressed as
\begin{equation}
    H = \sum_{\langle i,j \rangle}(J_x S_i^x S_j^x + J_y S_i^y S_j^y + J_z S_i^z S_j^z) + \sum_i h_z S_i^z ,
\end{equation}
with $\langle i,j \rangle$ depends on the specific lattice structure, $J_x,J_y,J_z$ describe the spin coupling strength respectively in the $x$,$y$,$z$ directions and $h_z$ is the magnetic field applied along the $z$ direction.  Classical approaches involving Bethe ansatz and thermal limitations~\cite{Bonechi_1992,faddeev1996algebraic,2001BrJPhR} have been developed and the model can be exactly solved. Besides, variational methods, in particular, VQE provides an alternative practical approach.  

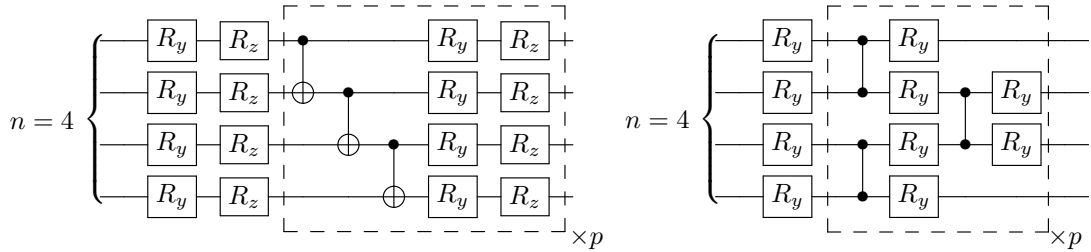
\begin{figure}[h]
\centering
\begin{subfigure}{.45\textwidth}
\centering
$$\Qcircuit @C=.9em @R=.4em{
&\qw &\gate{R_y} &\gate{R_z} &\ctrl{1} &\qw &\qw &\gate{R_y} &\gate{R_z} & \qw \\ 
&\qw &\gate{R_y} &\gate{R_z} &\targ &\ctrl{1} &\qw &\gate{R_y} &\gate{R_z} & \qw \\
&\qw &\gate{R_y} &\gate{R_z} &\qw &\targ &\ctrl{1} &\gate{R_y} &\gate{R_z} & \qw \\
&\qw &\gate{R_y} &\gate{R_z} &\qw &\qw &\targ &\gate{R_y} &\gate{R_z} & \qw \inputgroupv{1}{4}{0.4em}{3em}{n=4 \quad} \gategroup{1}{5}{4}{9}{1.2em}{--}\\
& & & & & & & &  & \\
& & & & & & & &  & \quad\times p
}
$$
\end{subfigure}
\begin{subfigure}{.45\textwidth}
\centering
$$\Qcircuit @C=.9em @R=.4em{
&\qw &\gate{R_y} &\qw &\ctrl{1} &\gate{R_y} &\qw      &\qw        &\qw &\qw    \\
&\qw &\gate{R_y} &\qw &\ctrl{0}   &\gate{R_y} &\ctrl{1} &\gate{R_y} &\qw &\qw     \\
&\qw &\gate{R_y} &\qw &\ctrl{1} &\gate{R_y} &\ctrl{0}   &\gate{R_y} &\qw &\qw    \\
&\qw &\gate{R_y} &\qw &\ctrl{0}  &\gate{R_y} &\qw &\qw   &\qw  &\qw \inputgroupv{1}{4}{0.4em}{3em}{n=4 \quad}
\gategroup{1}{5}{4}{8}{2.4em}{--} \\
& & & & & & & &  & \\
& & & & & & & & \times p & 
}$$     
\end{subfigure}
\caption{Four-qubit templates for the experimental ansatz (with $p$ repeated circuit block) diagrams of the \textbf{left}: Hardward Efficient ansatz (HEA) and the \textbf{right}: Alternating Layered ansatz (ALT).}
\label{fig:alt}
\end{figure}

For no-noise numerical experiments, we evaluate the performance of constant-depth VQE on randomly generated Hamiltonians with shuffled spin coupling strengths. Specifically, we consider the 1D model with adjacent interactions and randomly assign $J_{xyz}$ coefficients in the interval $[-1,1]$ while ignoring magnetic fields involving periodic boundary conditions. Furthermore, we utilise two commonly used parameterised quantum circuits or ansatzes namely hardware efficient ansatz (HEA)~\cite{Kandala_2017} and alternating layered ansatz (ALT)~\cite{Cerezo_2021}, as illustrated in Fig.~\ref{fig:alt}.  And for all simulations, we randomly initialise circuit parameters and perform 200 optimisation iterations using Adam optimiser with 0.1 learning rate.

\subsubsection*{Algorithm effectiveness via 1D Heisenberg model}
\label{subsec:overlap_amp}

\begin{figure}[htb]
    \centering
    \includegraphics[width=1.0\textwidth]{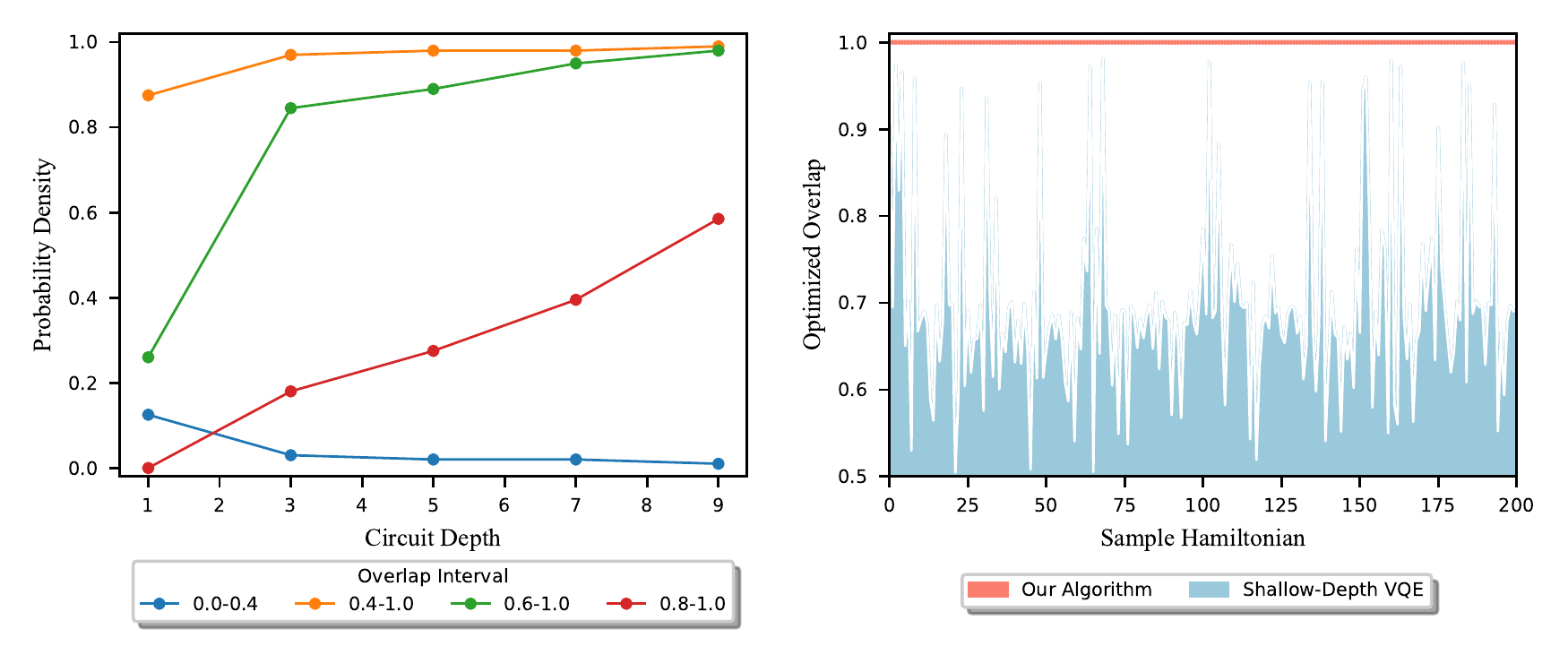}
    \caption{The left figure shows the overlap range prepared with ALT ansatz depth $[1,3,4,5,7]$ for 200 8-qubit Heisenberg Hamiltonians. The purple, blue, brown and orange lines indicate the intervals $(0.0-0.4)$, $(0.4-1.0)$, $(0.6-1.0)$ and $(0.8-1.0)$ in which the sampled probability of overlaps that prepared shallow-depth VQE may fall. The right figure shows the overlap prepared by depth-3 VQE with HEA ansatz and our algorithm. The blue zone represents overlap generated by shallow-depth VQE, while the red portion represents overlap produced by our method.}
    \label{fig:heisen_d_change_trend}
\end{figure}

First, we investigate the performance of initial state preparation as the circuit depth increases. In our simulation, we randomly sample 200 $8$-qubit Heisenberg Hamiltonians to demonstrate the initial state preparation by using ALT ansatz with circuit depth $d \in \{1,3,5,7,9\}$.  As shown in Fig.~\ref{fig:heisen_d_change_trend}, we categorise the trained overlaps into ranges $(0-0.4)$, $(0.4-1.0)$, $(0.6-1.0)$, and $(0.8-1.0)$ and designate them with purple, blue, brown and orange lines, respectively.  We observe that as the circuit depth increases, the probability of $(0.6-1.0)$ interval rises from $26\%$ to $98\%$ and $(0.8-1.0)$ overlap interval rises from $0$ to $58.5\%$, while $(0-0.4)$ interval drops from $12.5\%$ to $1\%$.  We also find that VQE with circuit depth can generate the initial state within $(0.4-1.0)$ overlap range with at least $97\%$ confidence.  This finding indicates that depth-3 ansatz can reliably prepare the initial state {for the ground state preparation of the 1D Heisenberg model} and a deeper-depth circuit may further boost the overlap.  As depth-3 ansatz is already capable of achieving satisfactory performance, we choose it for additional investigations.   

We then demonstrate the validity of our ground state preparation algorithms by applying the depth-3 VQE for preparing the initial state.  The experimental results are shown in Fig.~\ref{fig:heisen_d_change_trend}.  There are 200 8-qubit Heisenberg Hamiltonians being sampled, and HEA ansatz with circuit depth 3 is used.  The blue region depicts the overlap generated by a randomly initialised depth-3 VQE, while the red line represents the overlap boosted by our algorithm given the  initial states prepared before.  We can see that the shallow-depth VQE provides an initial state with a minimum overlap of 0.5 with the ground state. Then, by inputting the optimised state into our program, we can effectively prepare the ground state with a fidelity error of less than $10^{-4}$ regarding the true ground state. For remark, we use the Hamiltonian evolution operator $U=e^{iH}$ and set the estimation accuracy as $1/3$ of the spectral gap. 

\begin{figure}[htb]
    \centering
    \includegraphics[width=1.02\textwidth]{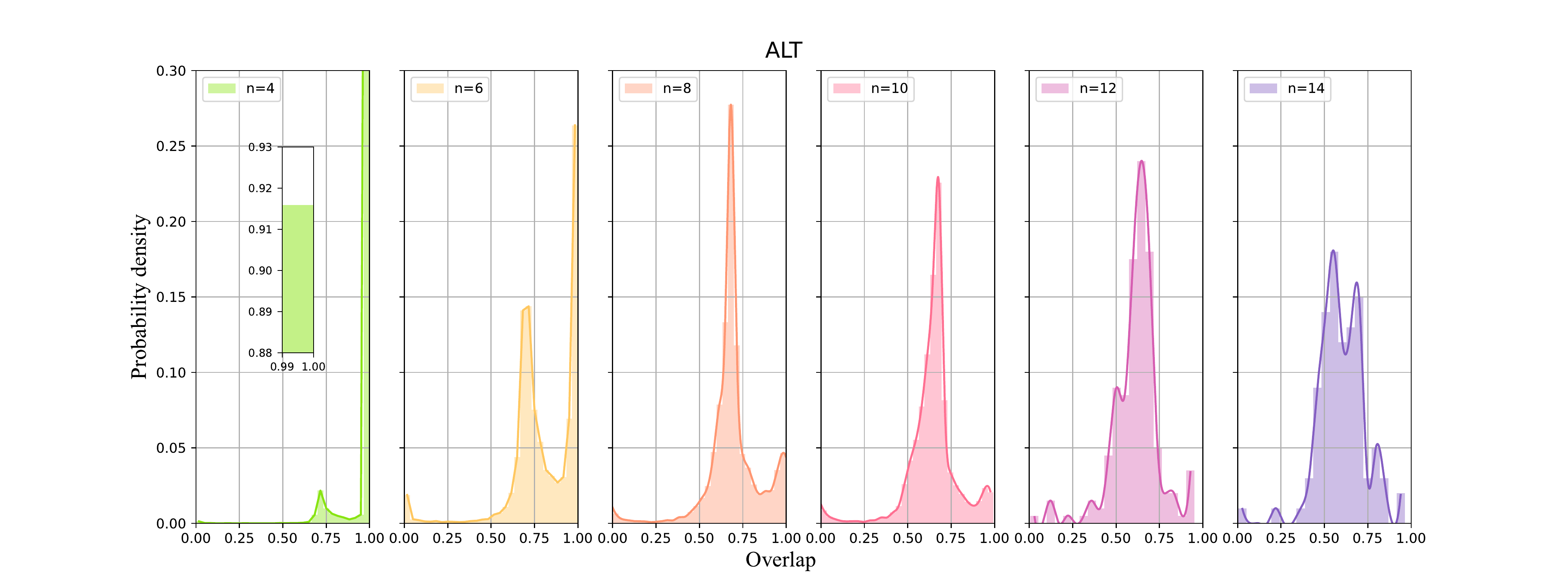} \\
    \includegraphics[width=1.02\textwidth]{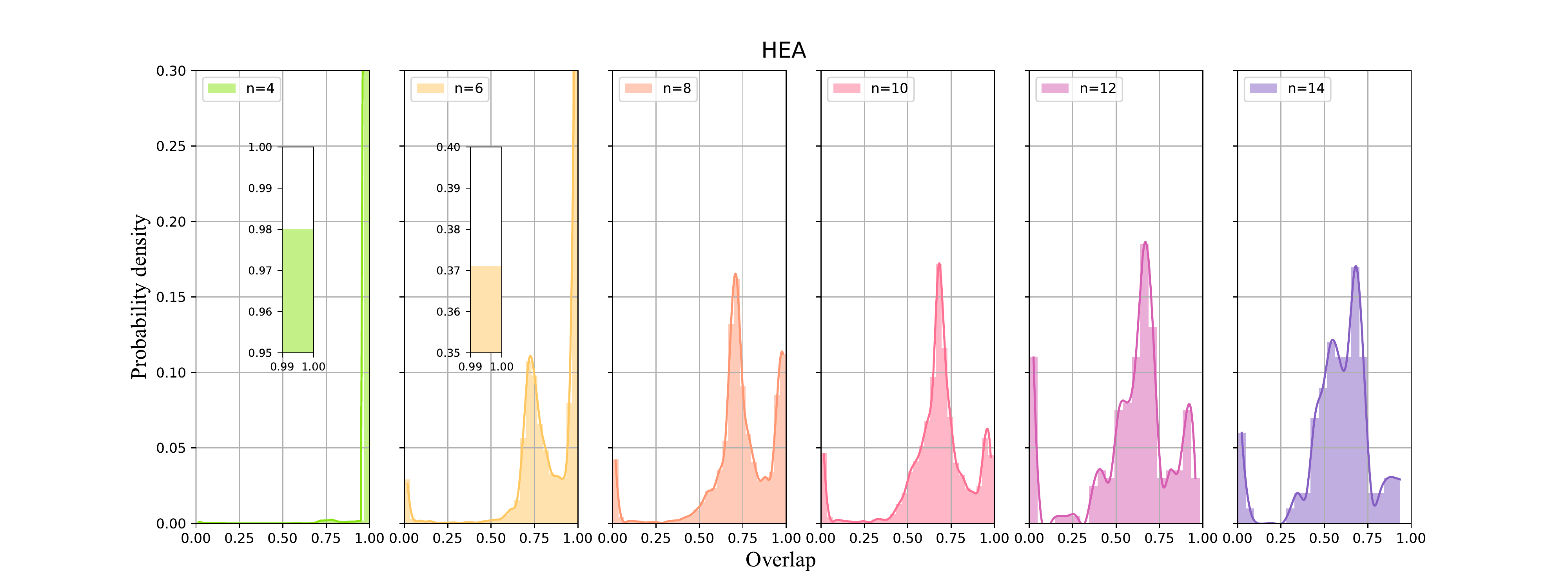}
    \caption{The frequency of overlap prepared by using depth-3 VQE ansatz of ALT (up) and HEA (down) for varied system sizes. The x-axis labels the overlap regarding the randomly generated Heisenberg Hamiltonians' ground states. For every system with a size less than or equal to 10 qubits, $10000$ random Hamiltonians are sampled, and for qubit sizes 12 and 14, $200$ and $100$ random Hamiltonians are sampled, respectively. The result shows the efficacy of depth-3 VQE of two ansatz templates in preparing initial states with an overlap greater than 0.4 for the intermediate-scaled Heisenberg models.}
    \label{fig:heisen_d3_vqe}
\end{figure}

Previous two experiments demonstrate the depth-3 VQE has satisfactory performance in the 8-qubit Heisenberg model.  To further understand the scalability of shallow-depth VQE in other system sizes, we conduct extensive noiseless numerical simulations for system sizes up to 14. Specifically, we sample 10000 randomly generated Heisenberg Hamiltonians for each qubit size between 4 to 10, and sample 200 and 100 random Hamiltonians for qubit sizes 12 and 14. As a comparison, ALT and HEA ansatz are both used. The experimental results are depicted in Fig.~\ref{fig:heisen_d3_vqe}.

According to the numerical results, the shallow-depth VQE is shown to be effective for Heisenberg Hamiltonians. The majority of the results indicate that the prepared state can have a greater overlap with the ground state than 0.4, making it an acceptable initial state for our algorithm. 
Also, as the system size of Heisenberg Hamiltonian increases, the prepared states have a trend that approximately converged to an overlap between 0.5 and 0.75 and the full width at half maximum of the sampled probability distribution expands, implying that, with a randomly initialised shallow-depth parameterised quantum circuit, the optimised state can be prepared with substantial overlap to the ground state with high probability. 
{Particularly, our results show that the technical concern from \cite{Ge2019,Dong2022,Lin2019a} can be fulfilled by a shallow-depth VQE, indicating the foreseeable possibility of realising the warm-start applications on large systems.}

Exceptionally, the experiments show that the overlap is less than 0.1 for a small number of Hamiltonians.  We alternatively utilise the Gibbs-VQE~\cite{Wang2020} method. 
In comparison to the shallow-depth VQE, Gibbs-VQE is less prone to prepare state with small overlap and has a more steady performance.  More discussions can be found in Appendix.~\ref{subsec:gibbs_vqe}.  Furthermore, due to the computational limitations, we are limited to observe the phenomenon up to the 14-qubit Heisenberg model. To further study the effectiveness of shallow-depth VQE, the Ising model is considered with system sizes greater than 20. A detailed discussion can be found in Appendix.~\ref{sec:ising_model}. 
Finally, barren plateaus \cite{McClean_2018} is a significant obstacle to the efficient use of VQE.  It refers to the gradient of the cost function vanishing exponentially with the number of qubits for a randomly initialised PQC with sufficient depth.  We show that shallow-depth VQE will not be affected by barren plateaus and is scalable to large-scale models, which can be referred to Appendix.~\ref{subsec:gradient_test}. 

\section{Application to chemical molecules} \label{sec:molecule}

\begin{figure}[t]
    \centering
    \includegraphics[width=0.75\linewidth]{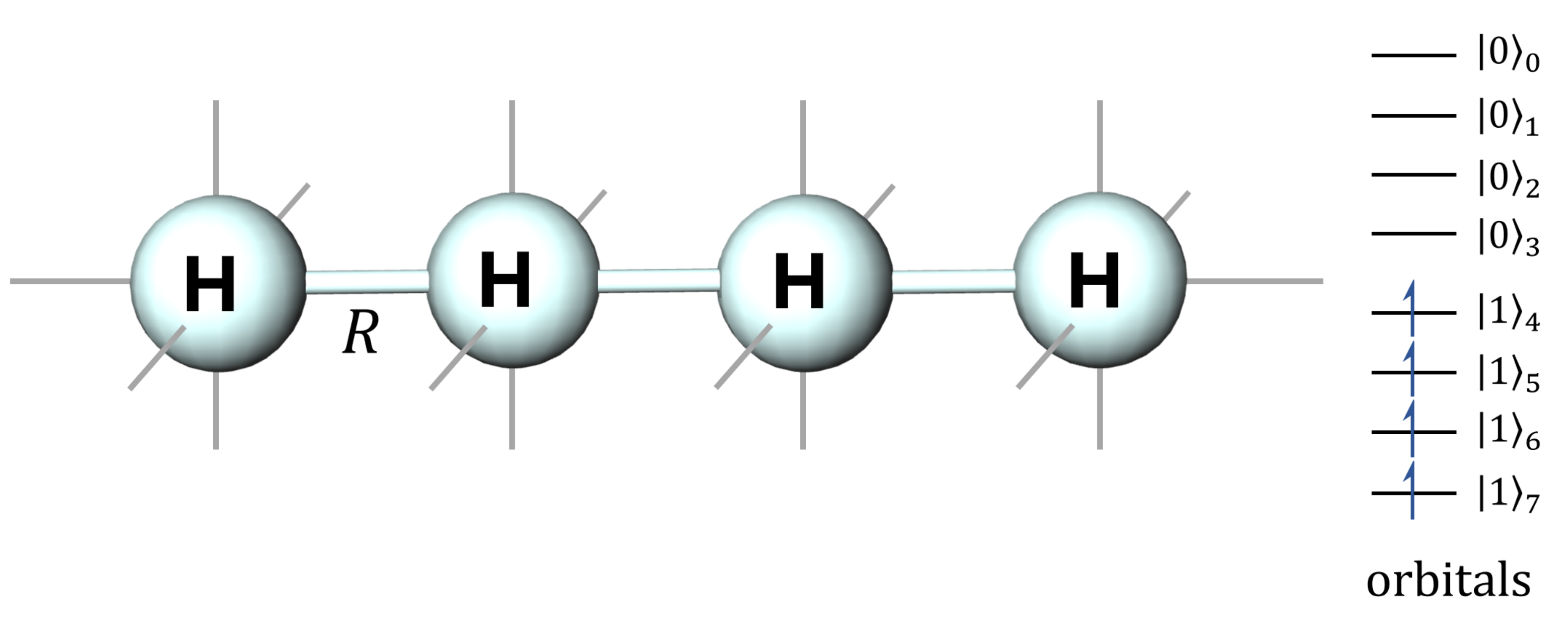}
    \caption{The molecular geometry of the hydrogen chain $H_4$ with $4$ atoms with equal spacing $R$. The right shows the qubit representation of the spin orbitals, and the subscripts label the qubit indices in the quantum circuit. The four spin-up orbitals are in the $\ket{1}$ state while no spin-down orbitals are occupied, representing the HF initial state of $H_4$.}
    \label{fig:hydrogen chain}
\end{figure}

Previous results have demonstrated the effectiveness of our algorithms for solving the Heisenberg model under intermediate-size situations and have the potential for large-size applications. Nevertheless, the systematic analysis of the ground state preparation performances with shallow-depth warm-start stays ambiguous for other physical models. {In this section, we aim to extend the usage of our algorithm into the quantum chemical regime by numerically demonstrating the warm-start performance on handling the small molecular model while taking the well-known Hartree-Fock (HF) state as a standard.}  

To our best knowledge, both VQE and HF state methods own abilities to prepare the high-accurate ground states~\cite{Dong_2021,Lin2019a,Wecker_2015} that can be used in quantum chemistry calculations for determining molecular geometries, and ground energies~\cite{bartlett1994applications,Peruzzo2014,Kandala_2017}. In the HF approximation, electrons are treated independently and indistinguishably~\cite{Omar_2005} that interact with nuclei potential and an electronic mean field~\cite{Rolf1977}. Quantum chemistry assumes discrete molecular orbitals determining the wave function of each electron appearing in the molecule. General molecular Hamiltonian fermionic configuration can be represented in a second quantisation form,
\begin{equation}\label{eq:mole ham}
    H = \sum_{pq} h_{pq} c_p^\dag c_q + \frac{1}{2}\sum_{pqrs} h_{pqrs} c_p^\dag c_q^\dag c_r c_s,
\end{equation}
where $c_p^\dag, c_p$ are the creation and annihilation operators, respectively. The coefficients $h_{pq}$ and $h_{pqrs}$ denote the one- and two-electron Coulomb integrals~\cite{Fermann2020}. In the quantum computing regime, the molecular electronic wave function can be represented in the occupation number basis that each qubit defines a spin-orbital. By adding one fermion (electron) into the system, each orbital can be in an occupied state $\ket{1}$ or a non-occupied state $\ket{0}$ represented in the computational basis. The single Slater-determinant state of $N$ orbitals can be defined as,
\begin{equation}
    \ket{\psi_\theta} = U_\theta\ket{\eta},\quad U_\theta = \exp{\left(\sum_{j,k=1}^N \theta_{jk} c_j^\dag c_k\right)}
\end{equation}
where $\ket{\eta} = c_\eta^\dag \cdots c_1^\dag \ket{0}$ in the core orbital basis. The Hartree-Fock state is the Slater-determinant having the lowest energy. Preparing the HF state classically requires a large amount of computational resource~\cite{google2020}. Even with quantum technologies~\cite{Tubman2018,Kivlichan2018}, one have to resolve the ansatz design for finding a basis to achieve relatively low circuit complexity~\cite{Babbush2018}. Besides, the extreme number of terms in the molecular Hamiltonian, writing in molecular orbital basis, could raise scalable challenging to simulate and measure which requires factorisation strategies~\cite{google2020,Berry2019qubitizationof}. 

\begin{figure}[t]
    \centering
    \subfloat[\centering H4 Bond distance 0.5]{{\includegraphics[width=0.45\textwidth]{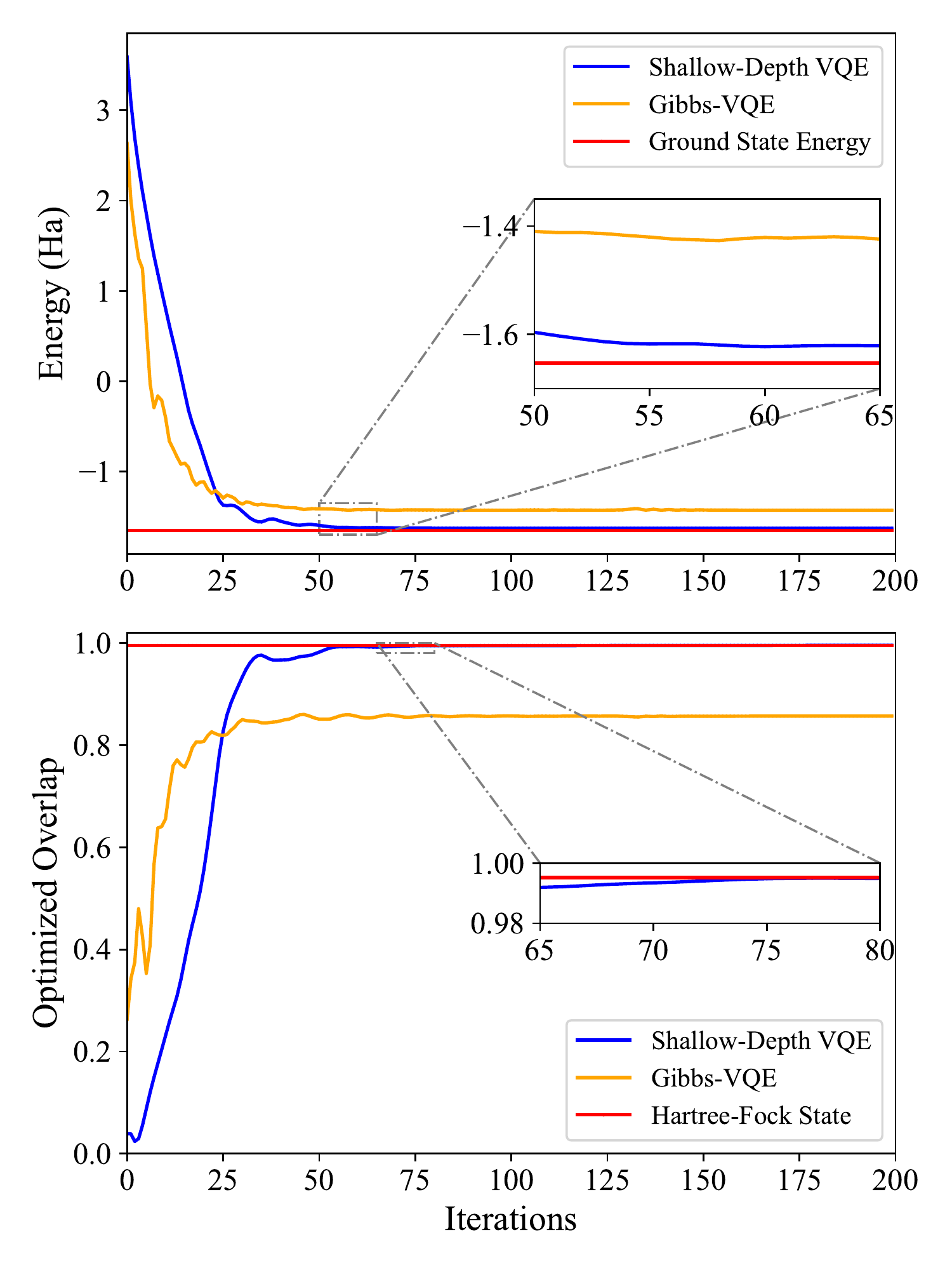} }}%
    \qquad
    \subfloat[\centering H4 Bond distance 3.0]{{\includegraphics[width=0.45\textwidth]{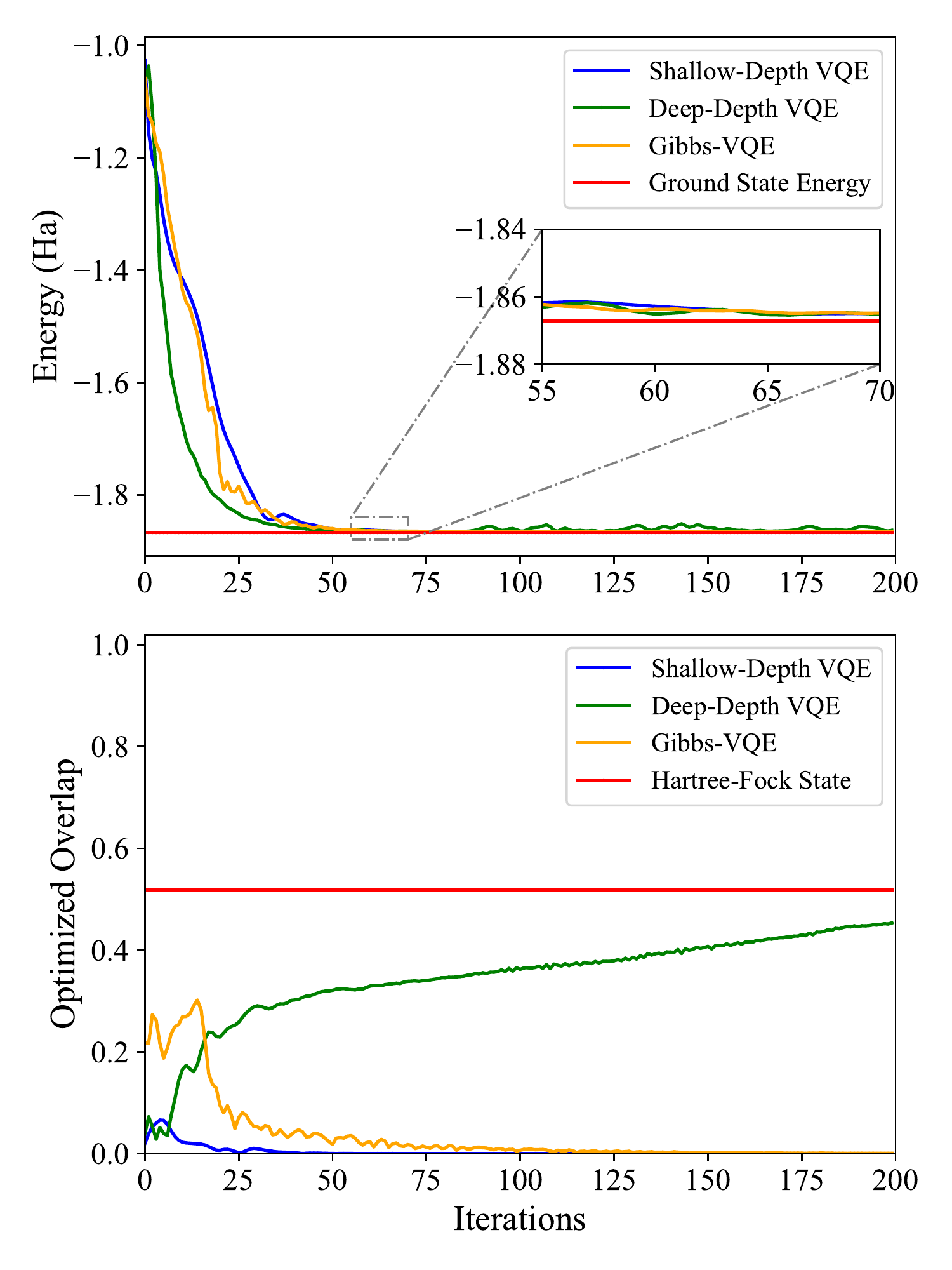} }}%
    \caption{Figures showing the ground state energy estimation and ground state preparation via shallow-depth (depth-3) VQE (blue), deep-depth (depth-50) VQE (green) and Gibbs-VQE (depth-3) (orange) on $H_4$ chain with bond distance $R$ equals to $0.5$ (a) and $3.0$ (b). The above diagrams illustrate the energy variations through the optimisation iterations, while the red lines represent the exact values. The bottom shows the corresponding prepared state overlaps, while the red shows the HF states' overlaps.}%
    \label{fig:VQE vs HF state}%
\end{figure}

We specifically consider the 1D hydrogen chain of $4$ atoms at fixed equal spacing $R$ containing $4$ electrons. Each hydrogen atom has one molecular orbital involving two spin orbitals that, in total, demands $8$ qubits for VQE. We choose hydrogen chain $H_4$ as the objective since its simplicity matches our current computational resource and its historical meaning as a benchmark for quantum chemistry~\cite{Motta2017}. We use \texttt{openfermion} built-in \texttt{Psi4} kernel to compute the fermionic Hamiltonian components~\eqref{eq:mole ham} of $h_{pq}$ and $h_{pqrs}$ via self-consistent field method using the STO-3G basis. We perform the simulations with $R=0.5$ and $3.0$ a.u. corresponding to the strong and weak interaction regimes, respectively. Before running the VQE, we need to map the fermionic Hamiltonian into the qubit representation via the Jordan-Wigner transformation~\cite{Jacob2012}. During all simulations, the VQE ansatz is chosen to be the HEA optimised in $200$ iterations of $0.1$ fixed learning rate by the Adam optimiser.

\subsubsection*{Comparing VQE and HF initial state on hydrogen chain}
 
Results have been illustrated in Fig.~\ref{fig:VQE vs HF state}. The top two diagrams illustrate the energy curves while the bottom shows the corresponding output states' overlap regarding the exact ground states. The decreasing energy curves' observation in the left plot matches the growth in the optimised overlap for all the cases of shallow-depth VQE (blue) and HF state (red) in (a) bottom. As the energy converges to around $-1.65$ Ha in (a) top, both the two schemes reach an overlap of almost one ($0.995$) in (a) bottom. Intuitively, bond distance $R=0.5$ a.u. creates strong electron-electron interaction making the mean field a reasonable approximation so that the HF state obtained a nearly perfect fidelity. This also explains the reduced overlap from the HF results reaching about $0.54$ when $R=3.0$ a.u. in (b) bottom since the approximation deviates. Counterintuitively, the VQE results, in blue, experience a distinct decline in the overlap though the corresponding energy values almost reach the ground of $-1.868$ Ha of error $10^{-3}$ compared to the exact diagonalisation results. {However, increasing the depth to $50$ for VQE ansatz (green) could gradually improve the overlap to $0.46$} within $200$ optimisation steps. We also include the Gibbs-VQE (yellow) stated in the previous sections as a reference scheme that showed similar behaviours from the shallow-depth VQE.

{We conclude that the shallow-depth VQE may be a scalable and preferable warm-starting strategy for molecular ground state preparation, showing great performance of generating high-overlap initial states for short bond distances.} 
Interestingly, for larger bond distances, both methods imply reduced initial overlap values. {As the mean field decoupled, the accuracy of the HF state decreases. Besides, the energy gaps between eigenstates shrunk traps the VQE into the local minimum~\cite{Wierichs_2020} since the close energy values carrying by the excited states could disturb the minimisation processes leading to a zero overlap by the orthogonality.} Deeper VQE circuit could mitigate the effect but with a slow growth of the overlap. Therefore, the ground energy estimation still performs well for governed by the variational principle, however, might be insufficient for the {efficient} initial state preparation. The fundamental pictures of the effect and the coping strategies remain open for extending the usage of our method on molecular models.

\section{Application to Fermi-Hubbard models}\label{sec:hubbard}
Researching the Fermi-Hubbard model is one of the fundamental priorities in condensed matter physics covering metal-insulator transitions and high-temperature superconductivity~\cite{Cade_2020,Dagotto1994}. However, the model involves a wide range of correlated electrons requiring multi-determinant calculations and numerous computational resources, hence impeding the classical exploration in the area. {Therefore, designing the quantum method for the model stays as a compelling application
for quantum computing. The potential of solving the 1D Hubbard model has been proposed by \citet{Dong2022} while only the VQE-typed algorithms have been examined via both numerical simulations and real devices~\cite{cheuk2016observation,Wecker_2015,Lin2019a}, as far as we know. In this section, we aim to provide a first numerical demonstration for the quantum ground state preparation scheme on the performance of predicting and estimating the physical properties of a classically challenging Fermi-Hubbard model other than commonly utilised VQE~\cite{Stanisic2021irm,google2020}. 
}

\begin{figure}[t]
    \centering
    \includegraphics[width=0.55\linewidth]{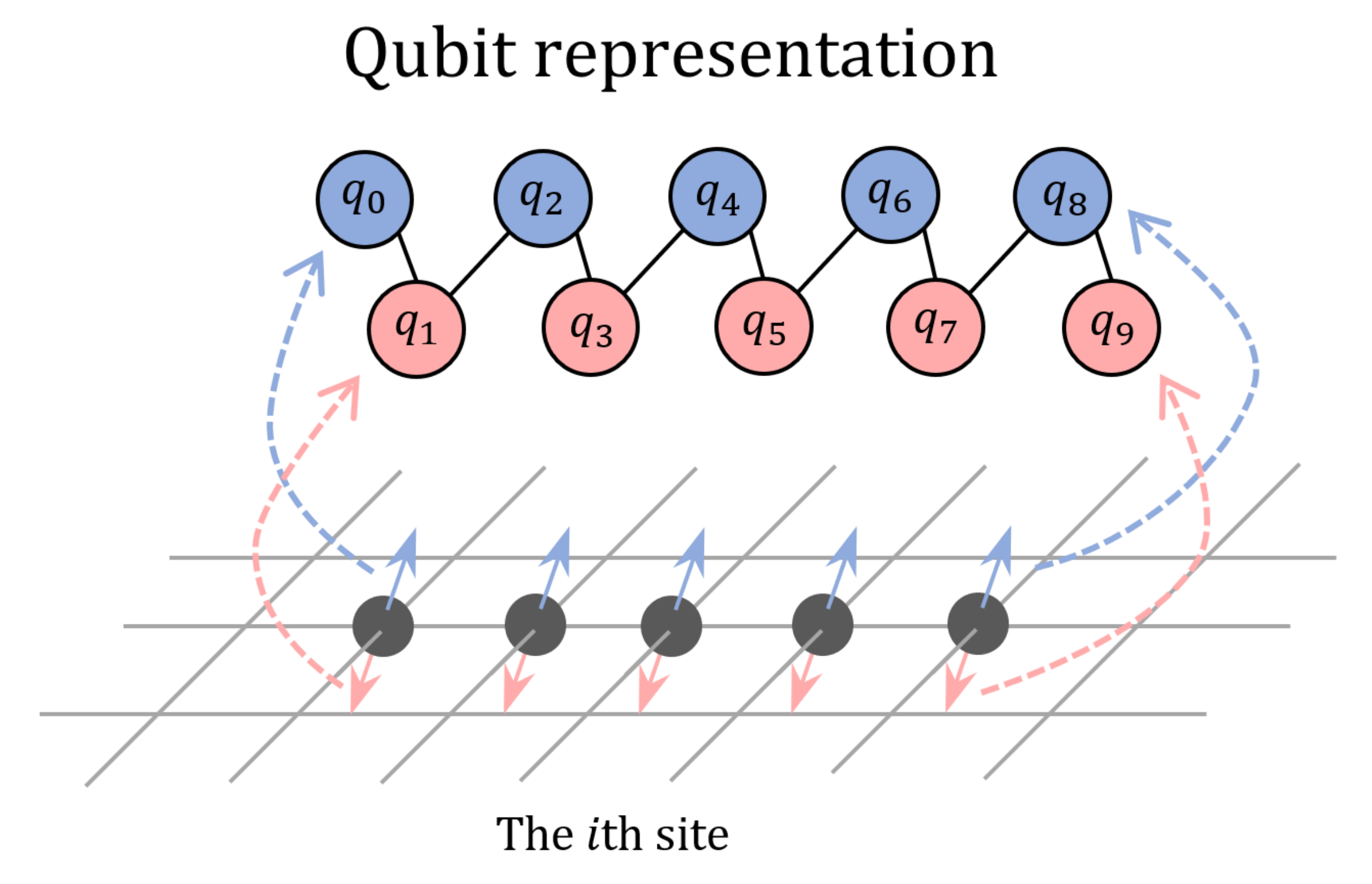}
    \caption{Qubit layout for implementing our $1\times 5$ Hubbard instances. The lower illustrate the model's physical site configuration where each site is associated with an index $1\leq i\leq 5$ labelling from left to right. Every site can hold a pair of spin-up and -down electrons which requires two qubits to encode all possible spin orbitals. Each qubit has perfect connections with the other qubits.}
    \label{fig:qubit representation of Hubbard}
\end{figure}

For an $n_x\times n_y$ square lattice physical system, for example, a metallic crystal, each lattice point, called a site, is assigned with an index. The Hubbard model Hamiltonian has a fermionic, second quantisation form,
\begin{equation}\label{eq:fermihubbard}
    H = -J\sum_{\langle i,j\rangle,\sigma}(a^\dag_{i\sigma}a_{j\sigma}+a^\dag_{j\sigma}a_{i\sigma}) + U\sum_i n_{i\uparrow}n_{i\downarrow} + H_{\text{local}},
\end{equation}
where $a^\dag_{i\sigma},a_{i\sigma}$ are fermionic creation and annihilation operators; $n_{i\sigma} = a^\dag_{i\sigma},a_{i\sigma}$, are the number operators; the notation $\langle i,j\rangle$ associates adjacent sites in the $n_x\times n_y$ rectangular lattice; $\sigma\in\{\uparrow,\downarrow\}$ labels the spin orbital. The first term in Eq.~\eqref{eq:fermihubbard} is the hopping term with $J$ being the tunnelling amplitude, and $U$ in the second term is the on-site Coulomb repulsion. The last term defines the local potential from nuclear-electron interaction. The local interaction term is determined by the Gaussian form~\cite{Wecker_2015},
\begin{equation}
    H_{\text{local}} = \sum_{j=1} \sum_{\nu = \uparrow,\downarrow} \epsilon_{j,\nu} n_{j, \nu}; \quad \epsilon_{j,\nu} = -\lambda_{\nu} e^{-\frac{1}{2}(j-m_{\nu})^2 / \sigma_\nu^2}.
\end{equation}
We consider an $1\times 5$ lattice model with $J=2,U=3$ and $\lambda_{\uparrow, \downarrow}=3, 0.1$, $m_{\uparrow, \downarrow} = 3, 3$. The standard deviation for both spin-up and -down potential is set to $1$. Such a setup guarantees a charge-spin symmetry around the centre site ($i=3$) of the entire system. 

As before, representing the Hubbard Hamiltonian on a quantum computer in the qubit representation requires fermionic encoding via efficient JW transformation. For a 1D Hubbard model with $N$ sites, based on the Pauli exclusion principle~\cite{Martin1951}, each site can contain at most a pair of spin-up and -down, two electrons. In total, there are $2N$ possible electronic orbitals represented by the same number of qubits which covers all possible orbitals for the system. The qubit layout of our five-qubit model has been shown in Fig.~\ref{fig:qubit representation of Hubbard} where all pairs of qubits are connected perfectly during simulations. We study the medium-scaled model having non-degenerate ground space. The fermionic Hamiltonian and qubitic transformations could be derived using \texttt{openfermion} and \texttt{paddle quantum} python libraries. The VQE uses an HEA circuit of depth $3$ and $50$ approaching realistic applications, and we choose the Adam optimiser (fixed learning rate $0.1$) as before training the circuit within $200$ iterations.

\subsubsection*{Evaluating physical quantities of Hubbard model}

\begin{figure}[t]
    \centering    \includegraphics[width=\linewidth]{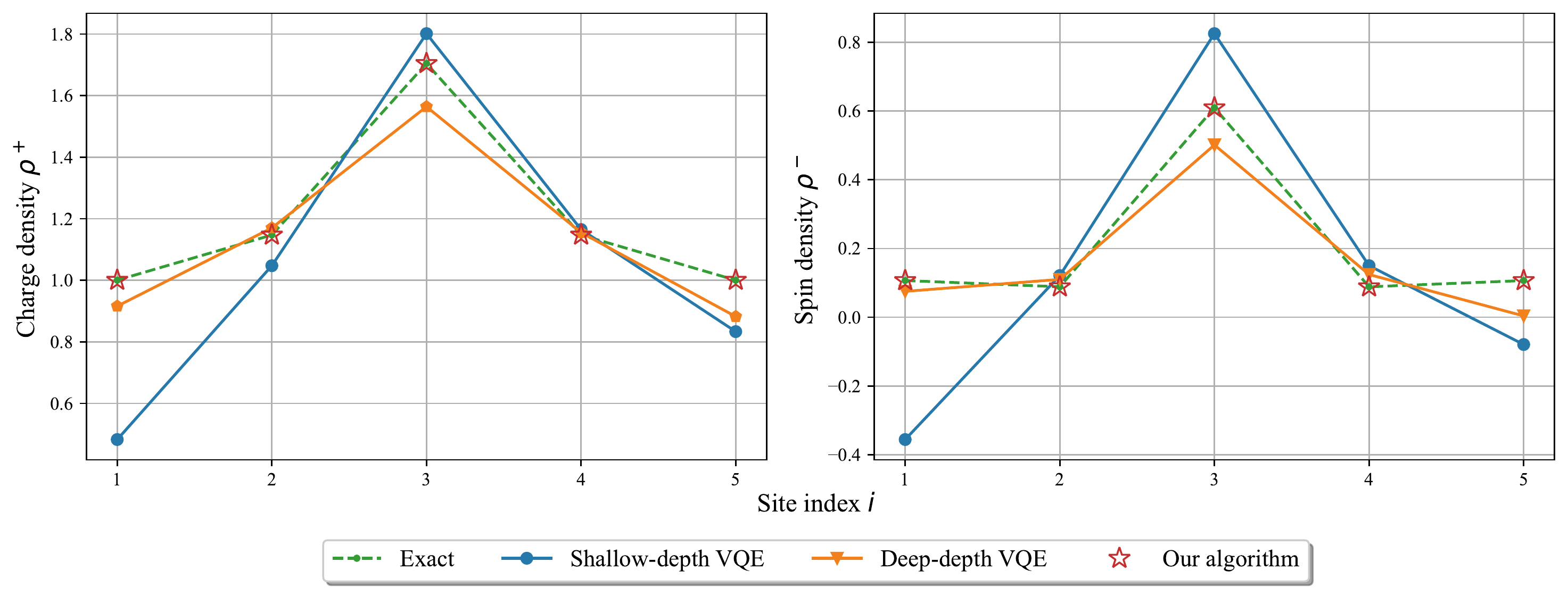}
    \caption{The (a) charge $\rho^+$ and (b) spin $\rho^-$ densities evaluated from different produced ground states of our stated Hubbard model ($5$ sites) by direct diagonalisation (green dashed), shallow-depth VQE (blue solid), deep-depth VQE (orange solid) and our algorithm (red star) methods. The shallow- and deep-depth VQE access HEA circuit with depth $3$ and $50$, respectively. In both cases, the learning rate is set to $0.1$, and the total number of iterations is set to $200$. The Adam optimiser powers all the training procedures.}
    \label{fig:charge-spin density of hubbard global gs}
\end{figure}

The first discussion is about predicting the charge-spin behaviours of the given solid system. We study the global ground state of the Hubbard model allowing the accessibility of every orbital with a sufficient number of electrons in the qubit representation. After deriving the ground state of the Hubbard Hamiltonian, we could compute the expectation value of different physical observables, for example, charge and spin densities defined as,
\begin{equation}
    \rho^{\pm}_j = \langle n_{j\uparrow}\rangle \pm \langle n_{j\downarrow}\rangle,
\end{equation}
where $j$ is the target site index. The densities could indicate the model's electronic and magnetic structures which further relate to the metal-metal bonds of given materials~\cite{Epsztein2018} and can be experimentally demonstrated via electron paramagnetic resonance experiments~\cite{Jacob_2012}, respectively. The density diagrams illustrate charge-spin distributions which could be used in designing solid structures of materials. 

We compare the evaluated ground state densities from different preparation schemes shown in Fig.~\ref{fig:charge-spin density of hubbard global gs}. As we could see both the densities show a centralized feature that matches our model configuration. The initial state of our algorithm is prepared via shallow-depth VQE deriving the corresponding density values (red stars) in the figure. A dramatic observation here shows our scheme outperforms the traditional VQE methods, in which the derived densities match the exact results while deep-depth VQE produces a visible gap regarding the true values (green dashed). Notice that a deeper circuit increases the VQE accuracy of both evaluated densities due to the better expressibility. However, such a circuit approximates 2-design which significantly drops the convergence speed of VQE. As a result, with the deep-depth circuit, VQE could only prepare a defective ground state and predict approximate densities (orange) within 200 iterations. 

Besides, our algorithm only requires a non-zero-overlap initial state coming from the shallow-depth VQE experiencing no trainability predicament. The searched state should theoretically be the exact ground state and predict exact density results. We show our method could derive high-precision charge-spin densities which can be used to verify the theoretical developments on Luttinger liquid and predict separation dynamics~\cite{arute2020observation}.

\begin{figure}[t]
    \centering    \includegraphics[width=0.55\linewidth]{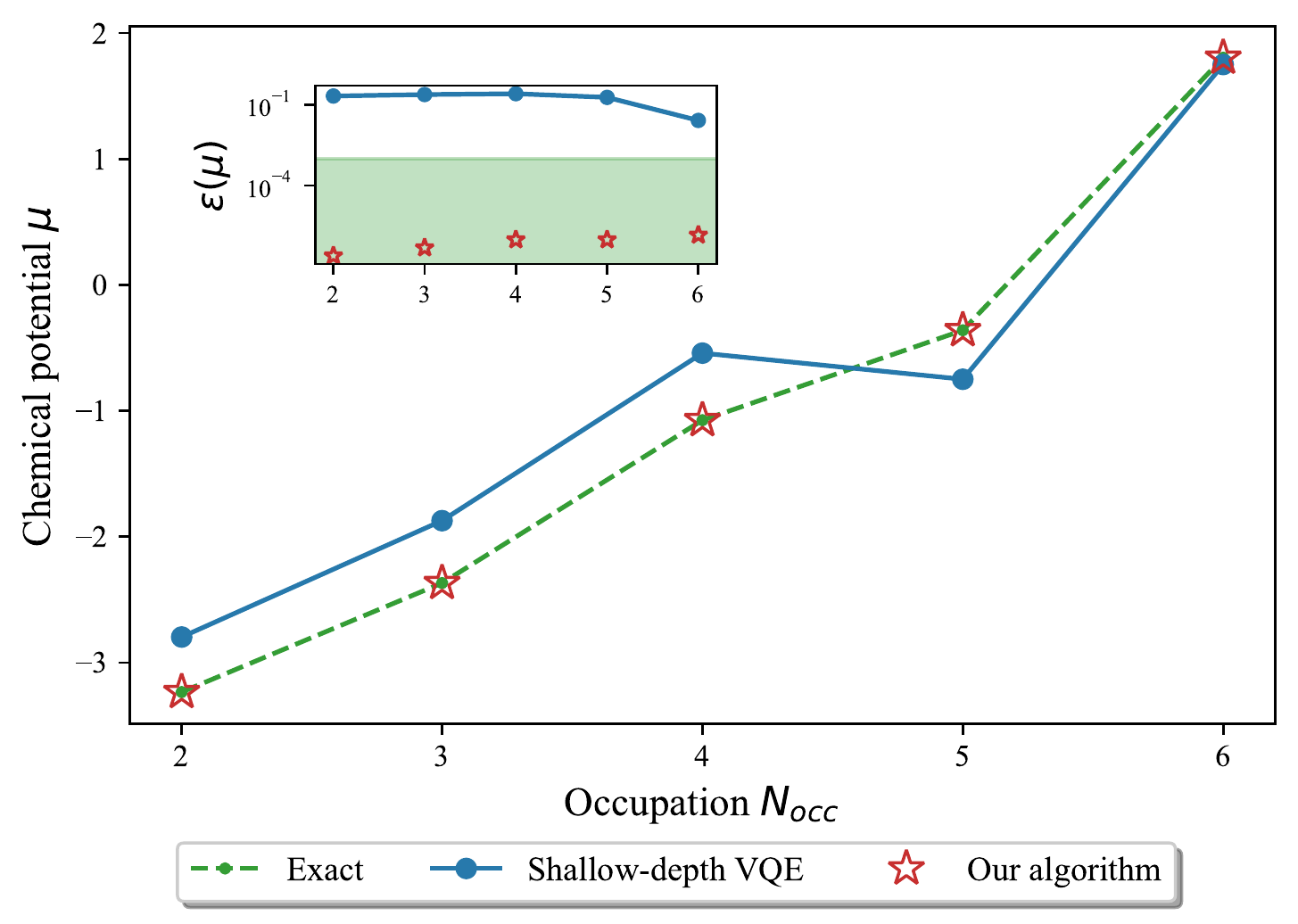}
    \caption{Chemical potential $\mu$ calculations of Hubbard ground states from different occupation numbers $N_{\text{occ}}$. The green dashed and blue solid lines illustrate the exact and shallow-depth VQE-prepared ground states, respectively. The stars label our algorithm's results by initialising the shallow-depth states. The subplot at the left corner indicates the corresponding potential values' absolute error $\varepsilon(\mu)$ (log-scale), where the green shaded area marks the points satisfying chemical accuracy regarding the FCI results.}
    \label{fig:chemical potential of hubbard}
\end{figure}

Apart from that, chemical potential is another physical quantity determining the energy variations during chemical and electronic interactions~\cite{Stanisic2021irm}. In the second quantisation, the potential fields are considered with discrete orbital states. Chemical potential describes the energy requirement by adding or subtracting an electron from the system. Such an electron would occupy an orbital state or leave a state empty, respectively. The value can be computed via subtracting the ground state energies of different occupation numbers (number of orbitals that are filled with particles), having the following expression,
\begin{equation}
    \mu(N_{\text{occ}}) = E(N_{\text{occ}}) - E(N_{\text{occ} }- 1).
\end{equation}
In the simulations, we derive the Hubbard Hamiltonian with different occupation numbers by classically applying fermionic projection maps to the previous global model by $\Tilde{H} = P_{N_{\text{occ}}}HP_{N_{\text{occ}}}$. The resulting $\Tilde{H}$ could have image space spanning by the standard basis elements with a fixed number of $\ket{1}$'s, and hence restrict the number of electrons in the system. {Such a projection can also be realised by so-called the number-preserving ansatz introduced in~\cite{Anselmetti_2021}.} Previous literature works on the deep HVA-based ansatzes ($30$ layers) and directly prepare the Hubbard ground states using VQE~\cite{Stanisic2021irm}. We instead, run the shallow-depth VQE on each $\Tilde{H}$ to prepare an initial state followed by our searching algorithm to locate the ground states of each occupation number {with constant scaled phase factor}. The corresponding energies can be simultaneously estimated from our algorithm. 

As from Fig.~\ref{fig:chemical potential of hubbard}, the algorithm could successfully extract the ground state from initial shallow-depth VQE's states. By zooming into the subplot illustrating the absolute error of the chemical potentials at each $N_{\text{occ}}$. Our estimated values obtain an absolute error of almost $10^{-5}$ which have all reached the chemical precision with respect to the FCI calculations via noiseless simulations in the green-shaded area. The observation significantly proves the effectiveness of our method on other high-accurate condensed matter calculations which inspires that our scheme could promote a variety of physical research and provide a new paradigm for studying and predicting hard-to-solve model behaviours from both classical and quantum horizons.

\section{Conclusion and outlook}\label{sec:conclusion}
In this paper, we provide quantum algorithms to prepare ground states and demonstrate the correctness and effectiveness through extensive numerical simulations on Heisenberg models and chemical molecules. We also apply our algorithms to estimate physical quantities of interest, such as charge and spin densities and chemical potential of Fermi-Hubbard models having no generally efficient classical methods, to our best knowledge. Only VQE and its variants have been applied and demonstrated with numerical and practical experiments from recent literature. A notable property of our algorithms is that the ground state can be prepared simultaneously with the ground state energy estimation, thus requiring no prior information on the ground state energy. {In contrast, existing ground state preparation algorithms involving quantum phase estimation have to estimate a proper ground state energy beforehand} \cite{Ge2019}, leading to indefinite practicability. For other works \cite{Lin2019a,Dong2022} based on QSVT and QET-U algorithms, a reasonable upper bound estimate of the ground state energy is pre-requested. Since such information is usually not known as priorities, one has to expensively estimate the ground state energy before preparing the state. {Another worth-noting point is that our algorithms only require to compute phase factors with a scaling $\widetilde{\mathcal{O}}(1)$ while QSVT and QET-U based algorithms require $\widetilde{\mathcal{O}}(\Delta^{-1})$, which means our algorithms have advantages in avoiding challenging phase factor evaluations when $\Delta$ is small}. 

Besides algorithmic improvements, we also contribute to showing that shallow-depth VQE can be a favourable method for warm-starting the ground state preparation of the intermediate-scaled many-body systems. In contrast, the previous works \cite{Ge2019,Dong2022,Lin2019a,Dong2022} all assume the state with a considerable overlap is available and can be reused.  
Our scheme provides a potential paradigm for using shallow-depth VQE to solve challenging physical models. By conducting simulations to a 10-qubit Hubbard model, our method outperforms the VQE-based evaluations of the charge and spin densities and the chemical potentials. Meanwhile, we efficiently and qualitatively predict the ground state density behaviours of the potential centralised model and derive extremely accurate chemical potentials of distinct occupations up to an absolute error of $10^{-5}$. Such evaluations have reached the chemical accuracy of the FCI results, which is foreseeable to own advantages to exploring high-temperature superconductivity and quantum chemistry problems. Our simulations have demonstrated the potential of our algorithm on solving the 1D Hubbard model and prepared for further research on the classical challenging 2D models~\cite{cheuk2016observation}.

{Ground state preparation is significant to the fundamental research in physics and chemistry. To establish a synthetic quantum solution for this task, there are still some remaining problems to resolve in future. Since our works can not offer suggestions on the depth of the warm-start ansatz, a theoretical guarantee of reaching the large initial state overlap using sufficient-depth ansatzes stays required.} 
Recall that the algorithms are based on the prior information on the spectral gap, {especially, when the spectral gap is small, it would be better to amplify it economically}. Hence, an efficient and effective method to compute the spectral gap {and even amplify the gap} is necessary for applying quantum computers in solving ground-state problems. 
{In addition to addressing many-body systems, we believe our techniques can also be used for combinatorial optimization problems e.g. the travelling salesperson problem.} Based on the results in this paper, we believe that {shallow variational warm-start} would find more applications in physics, chemistry, and quantum machine learning, which is worth further studying in theory and experiments.

\section*{Acknowledgement}
Y. W., C. Z., and M. J. contributed equally to this work. Part of this work was done when Y. W., C. Z., and M. J. were research interns at Baidu Research.  Y. W. acknowledges support from the Baidu-UTS AI Meets Quantum project.
The authors would like to thank Lloyd C.L. Hollenberg and Haokai Zhang for the valuable comments and Lei Zhang for the helpful discussion.

\bibliography{ref}

\clearpage

\appendix
\setcounter{subsection}{0}
\setcounter{table}{0}
\setcounter{figure}{0}

\vspace{2cm}
\onecolumngrid
\vspace{2cm}

\begin{center}
\textbf{
{\large{Supplementary Material for \\ Ground state preparation with shallow variational warm-start}}}
\end{center}

\numberwithin{equation}{section}
\renewcommand{\theproposition}{S\arabic{proposition}}
\renewcommand{\thedefinition}{S\arabic{definition}}
\renewcommand{\thefigure}{S\arabic{figure}}
\setcounter{equation}{0}
\setcounter{table}{0}
\setcounter{section}{0}
\setcounter{proposition}{0}
\setcounter{definition}{0}
\setcounter{figure}{0}


\section{Initial State Preparation with Gibbs-VQE}
\label{subsec:gibbs_vqe}
The experiments in Fig.~\ref{fig:heisen_d3_vqe} show that it is unsuccessful for a small number of Hamiltonians.  This may be caused by the local minima problem, which we re-initialise all the circuit parameters and re-run the circuit again may significantly amplify the overlap.  To mitigate this issue, we consider applying variational Gibbs-state preparation\cite{Wang2020} (Gibbs-VQE) alternatively.

We begin by briefly reviewing the background of the Gibbs state and the protocol Gibbs-VQE. For a Hamiltonian with an $n$-qubit system, the Gibbs state $\rho_G$ at temperature $T$ is defined as follows:
\begin{equation}
    \rho_G = \frac{e^{-\beta H}}{\tr(e^{-\beta H})},
\end{equation}
where $-\beta H$ is the matrix exponential of matrix $-\beta H$. $\beta = \frac{1}{kT}$ is the inverse temperature of the system and $k$ is the Boltzmann's constant.  To prepare a Gibbs state we use the protocol as introduced in \cite{Wang2020}.  The main idea is to consider the variational principle of the system's free energy, where parameterised quantum circuits are used to minimize the free energy.  Specifically, we set the loss function as $L = L_1 + L_2 + L_3$, where
\begin{equation}
    L_1 = \tr{H_{\rho_B}}, \quad L_2 = 2\beta^{-1}\tr{\rho_B^2}, \quad L_3 = \frac{-\beta^{-1}(\tr{\rho_B^2}+3)}{2}. 
\end{equation}
       
\begin{figure}[htb]
    \centering
    \includegraphics[width=0.6\textwidth]{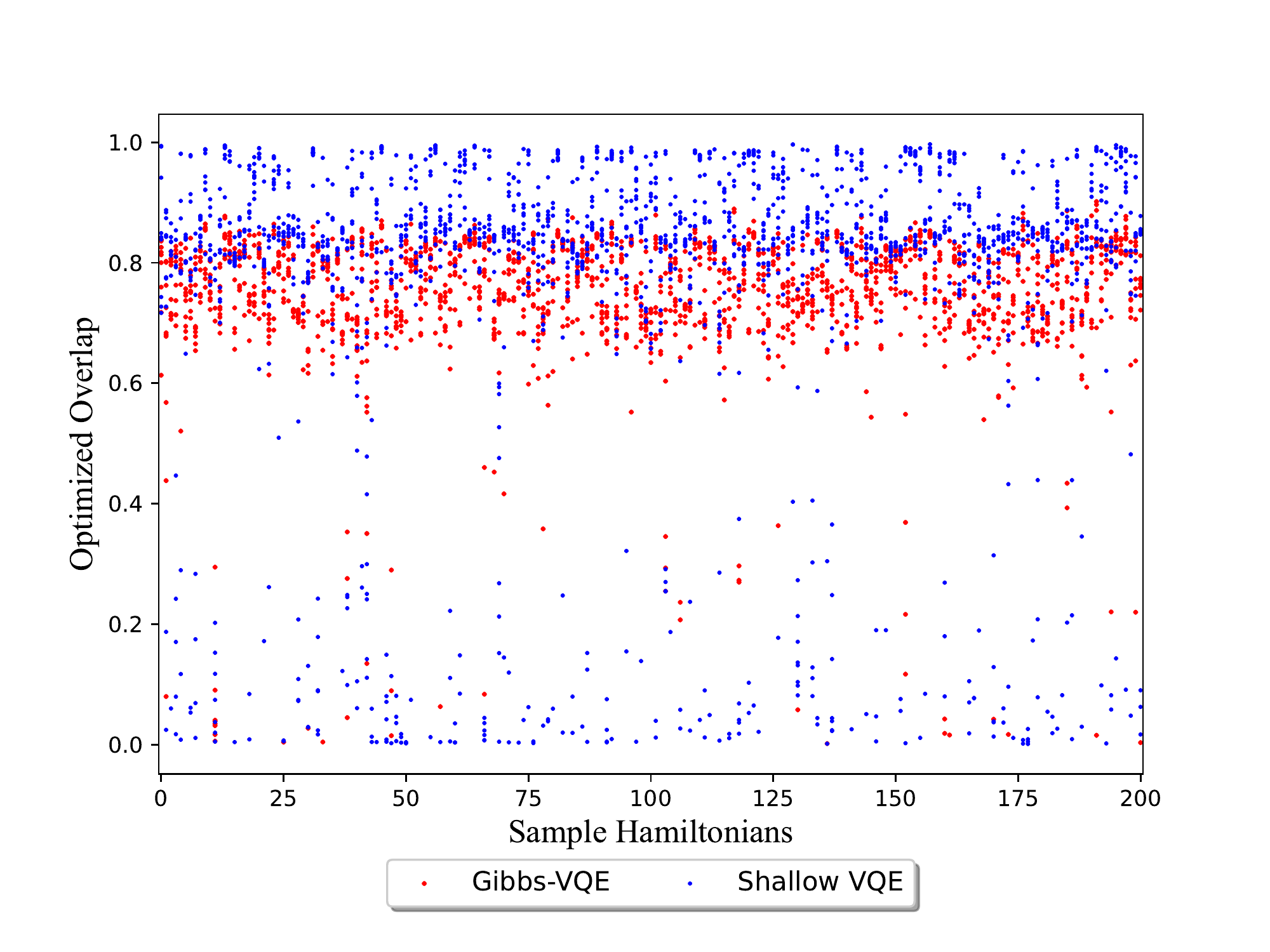} 
    \caption{The optimised overlap using shallow-depth VQE and shallow-depth Gibbs-VQE. For each sampled Hamiltonian, 10 optimisations are performed to determine the overlap between the optimised and ground states. The blue circle indicates the overlap prepared with shallow-depth VQE and the red diamond represents the case with the Gibbs-VQE.}
    \label{fig:overlap_gibbs_shallow}
\end{figure}

To compare the performance of Gibbs-VQE and shallow-depth VQE, 200 random Hamiltonians are sampled and 10 random initialisations of circuit parameters are performed for each Hamiltonian. For both simulations of two methods, we employ depth-3 HEA ansatz and 200 Adam optimiser optimisation iterations.  The experiment results are shown in Fig.~\ref{fig:overlap_gibbs_shallow}.  The blue circle indicates the overlap prepared by shallow-depth VQE, and the red diamond indicates the overlap prepared by shallow-depth Gibbs-VQE.  Simulation results indicate that Gibbs-VQE has a more steady performance than shallow-depth VQE and is less prone to become stuck in local minima.  It can be seen that $\sim11\%$ number of shallow-VQE prepare the overlap less than 0.4, whereas, for Gibbs-VQE, the ratio reduces to $\sim2\%$.  Thus, Gibbs-VQE is superior for solving physical and chemical problems because it generates initial states with reliable overlap.

\section{Initial State Preparation with Tensor Product Ansatz}
\label{sec:ising_model}
The process that initialises an approximating state for a quantum algorithm via classical methods is usually called a \textit{warm-start} quantum method. By inheriting the initial state, quantum algorithm could require fewer resources for solving the problem. Famous result, for example, warm-start QAOA~\cite{Egger_2021} combining Goemans-Williamson (GW) randomized rounding algorithm, guarantees the performance of solving the max-cut problem with $\cO(n)$ qubits and low-depth quantum circuits.  

From the previous section~\ref{sec:heisenberg}, we observe the efficiency of initial state preparation on Heisenberg Hamiltonians through shallow-depth VQE.  Due to the computational limitations, we are limited to observe the phenomenon on Hamiltonians up to $14$-qubit.  To further investigate the efficacy of shallow-depth VQE, we use the tensor product ansatz to examine the max-cut problem.  Specifically, the max-cut problem can be naturally mapped to the QUBO Hamiltonian. A general \textit{QUBO} Hamiltonian has an expression,
\begin{equation}\label{qubo_hamiltonian}
    H_{\text{QUBO}} = \frac{1}{2}\sum_{p\neq q} Z_p Z_q - \frac{n}{2}\mathbb{I}^{\otimes n}.
\end{equation}
This is closely related to the continuous relaxation of the max-cut problem which can be approximately solved using semidefinite programming and GW rounding. Such a technique generates the initial state for the problem Hamiltonian, which can be efficiently prepared by applying a single layer of $R_Y$ rotations. This indicates the possibility of the approximating ground state of specific Hamiltonian models via shallow-depth quantum circuits. Taking the max-cut problem as a starting point, we could have the following observation. 
\begin{lemma}\label{lemma:QUBO+RY}
    For a product state ansatz composed with only one layer of single-qubit $R_y$ rotation, s.t., $\ket{\psi(\theta)} = \bigotimes R_y (\theta_p)\ket{0}$, we have
\begin{equation}\label{eq:product ansatz cost for qubo}
    \braket{\psi(\theta)}{H|\psi(\theta)} = \frac{1}{2}\sum_{p\neq q} \braket{\psi(\theta)}{Z_p Z_q|\psi(\theta)} - \frac{n}{2}\braket{\psi(\theta)}{\mathbb{I}^{\otimes n}|\psi(\theta)} = \frac{1}{2}\sum_{p\neq q} \cos{(\theta_p)}\cos{(\theta_q)} - \frac{n}{2}.
\end{equation}
\end{lemma}
As from above, we could work out the analytic gradient for any k-local Hamiltonian with the product state ansatz $\ket{\psi(\theta)}$. For QUBO case, setting cost $C(\bm{\theta}) = \bra{\psi(\theta)} H_{\text{QUBO}}\ket{\psi(\theta)}$, we could easily derive,
\begin{equation*}
    \partial_\alpha C = -\sin{(\theta_\alpha)}\sum_{p\neq \alpha}\cos{(\theta_p)}
\end{equation*}

\begin{figure}[h!]
    \centering
    \includegraphics[width=0.5\textwidth]{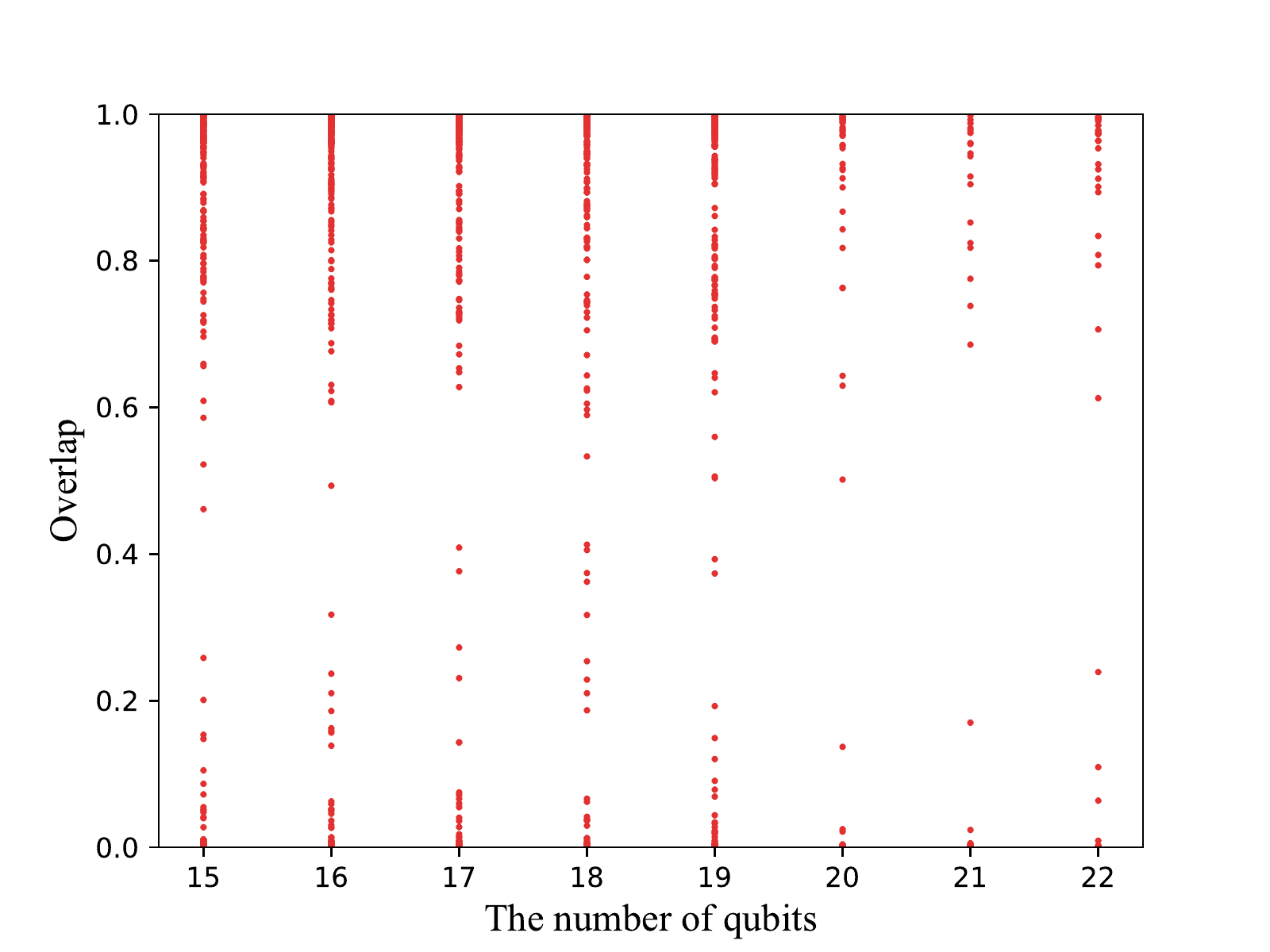}
    \caption{{The highest overlap between the optimised state and one of the ground states of different given QUBO Hamiltonian models (each model probably has degenerate ground state space) vs. the system sizes}. For the number of qubits between 15 to 19 and 20 to 22, we sample $500$ and $100$ random max-cut graphs' qubo Hamiltonians respectively. The true ground states of each model were first determined using brute force algorithm and stored in classical memory. Each model will then be processed in VQE with the tensor product ansatz and the optimisation was done w.r.t the analytic form Eq.~\eqref{eq:product ansatz cost for qubo} to produce an optimised state.}
    \label{fig:qubo_overlap}
\end{figure}

We then use QUBO Hamiltonian and the result in Lemma \ref{lemma:QUBO+RY} to further investigate the performance of shallow-depth VQE. With powerful classical machine learning techniques, we could classically derive the optimal rotation angles via minimizing the cost \eqref{eq:product ansatz cost for qubo}. The preparation of the approximating initial state is then implemented by applying the corresponding rotations on $\ket{0}$. In this experiment, we investigate QUBO Hamiltonians with a size of 15 to 22 qubits and calculate the overlap using a brute-force technique. The experiment results are shown in Fig. \ref{fig:qubo_overlap}.  

As we seen from the diagram, the x-axis represents the size of QUBO Hamiltonian and y-axis is the overlap between the initial state prepared by analytic form according to E.q. \ref{eq:product ansatz cost for qubo} and the ground states.  The probability of getting high overlap states regarding one of the ground states is slightly higher than trapping into the excited states (local minima).  The experiment trend suggests that shallow-depth VQE may continue to perform well on large-scale Hamiltonians.

\section{Numerical Simulations on the Gradient of Shallow-Depth Circuit}
\label{subsec:gradient_test}

The scalability is a significant obstacle to the efficient use of variational quantum eigensolver.  Numerous studies have demonstrated that the gradient of the cost function vanishes exponentially with the number of qubits for a randomly initialised PQC with sufficient depth~\cite{McClean_2018}.  This phenomenon is termed the barren plateaus (BP).  It has been shown that hardware efficient ansatz with deep depth leads to exponentially vanishing gradients~\cite{McClean_2018}. While Alternating Layered Ansatz, a specific structure of the HEA ansatz~\cite{Cerezo_2021}, has been demonstrated that when circuit depth is shallow, it does not exhibit the gradient vanishing problem~\cite{Nakaji_2021}.  As we explore using the HEA and ALT ansatz as an overlap initialiser, it is crucial to determine if these two ansatz with shallow depth suffer from the gradient vanishing issue.

\begin{figure}[ht]
    \centering
    \includegraphics[width=0.5\textwidth]{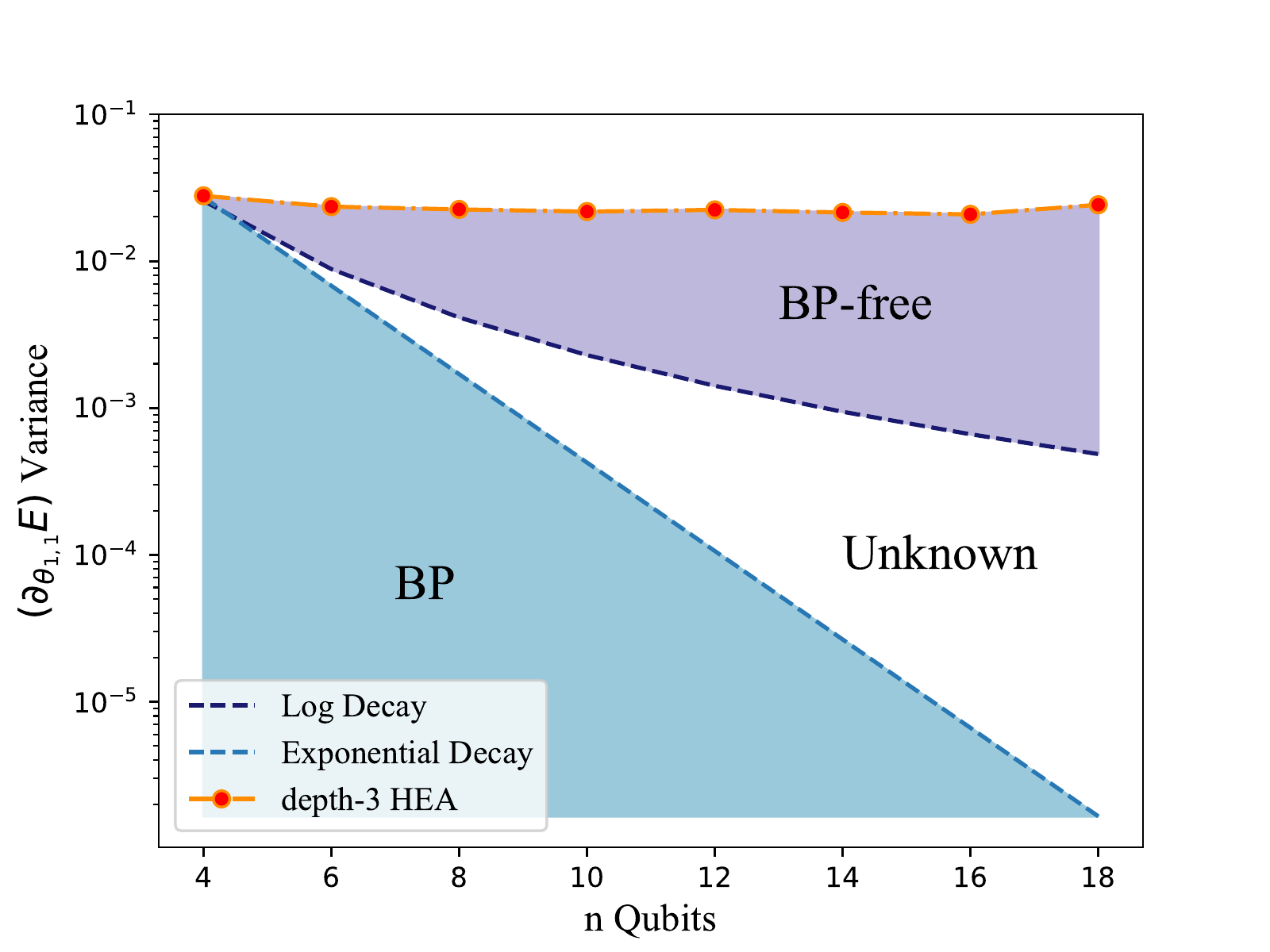}
    \caption{Semi-log plot displaying the sample variance of the gradient of a two-local Pauli term as a function of the number of qubits. The blue line represents exponential decay as a function of the number of qubits, the dark blue line represents the logarithmic decay and the orange line indicates that the shallow-depth VQE do not exhibit exponential decay.}
    \label{fig:gradient_shallowVQE}
\end{figure}

As shallow-depth ALT ansatz is BP-free, we choose to use the depth-3 HEA ansatz as an illustration and conduct numerical experiments to sample the variance of the gradient.  The experiment results are shown in Fig.~\ref{fig:gradient_shallowVQE}.  Similar to the experimental setting described in~\cite{McClean_2018}, we analyse the Pauli $ZZ$ operator acting on the first and second qubits such that $H=Z_1 Z_2$.  The gradient is then sampled on the first parameter $\theta_{1,1}$ by taking 1000 samples and the system size is evaluated up to 18 qubits. 

As shown in the figure, the orange line indicates the sample variance of the gradient of the shallow-depth VQE while the blue and dark blue lines depict exponential and logarithmic decay, respectively.  Additionally, we identify the blue area below the exponential decay line as the BP area and the purple area above the logarithmic decay as BP-free.  It is evident that the depth-3 HEA ansatz does not suffer from the gradient vanishing problem.  Therefore, it is scalable to use shallow-depth VQE for initial state preparation.


\end{document}

%% file: pretex.tex
\usepackage{algorithm,algorithmic,amsfonts,amsmath,amssymb,mathtools,mathrsfs,bbm,float}
\usepackage{graphicx,epic,eepic,epsfig,latexsym,verbatim,color}
\usepackage[shortlabels]{enumitem}
 
\usepackage{tikz}
\usetikzlibrary{chains}
\usetikzlibrary{fit}
\usepackage{pgflibraryarrows}		
\usepackage{pgflibrarysnakes}	

\usepackage{epsfig}
\usetikzlibrary{shapes.symbols,patterns} 
\usepackage{pgfplots}

\usepackage[strict]{changepage}
\usepackage{hyperref}
\hypersetup{colorlinks=true,citecolor=blue,linkcolor=blue,filecolor=blue,urlcolor=blue,breaklinks=true}

\usepackage[marginal]{footmisc}
\usepackage{url}
\usepackage{theorem}

\newtheorem{proposition}{Proposition}
\newtheorem{lemma}[proposition]{Lemma}

\newtheorem{theorem}[proposition]{Theorem}

\def\squareforqed{\hbox{\rlap{$\sqcap$}$\sqcup$}}
\def\qed{\ifmmode\squareforqed\else{\unskip\nobreak\hfil
\penalty50\hskip1em\null\nobreak\hfil\squareforqed
\parfillskip=0pt\finalhyphendemerits=0\endgraf}\fi}
\def\endenv{\ifmmode\;\else{\unskip\nobreak\hfil
\penalty50\hskip1em\null\nobreak\hfil\;
\parfillskip=0pt\finalhyphendemerits=0\endgraf}\fi}

\newcounter{remark}

\newcounter{example}

\mathchardef\ordinarycolon\mathcode`\:
\mathcode`\:=\string"8000
\def\vcentcolon{\mathrel{\mathop\ordinarycolon}}
\begingroup \catcode`\:=\active
  \lowercase{\endgroup
  \let :\vcentcolon
  }

\usepackage{cleveref}
\usepackage{graphicx}
\usepackage{xcolor}

\definecolor{darkblue}{RGB}{0,76,156}
\definecolor{darkkblue}{RGB}{0,0,153}
\definecolor{blue2}{RGB}{102,178,255}
\definecolor{darkred}{RGB}{195,0,0}

\RequirePackage[framemethod=default]{mdframed}
\newmdenv[skipabove=7pt,
skipbelow=7pt,
backgroundcolor=darkblue!15,
innerleftmargin=5pt,
innerrightmargin=5pt,
innertopmargin=5pt,
leftmargin=0cm,
rightmargin=0cm,
innerbottommargin=5pt,
linewidth=1pt]{tBox}

\newmdenv[skipabove=7pt,
skipbelow=7pt,
backgroundcolor=blue2!25,
innerleftmargin=5pt,
innerrightmargin=5pt,
innertopmargin=5pt,
leftmargin=0cm,
rightmargin=0cm,
innerbottommargin=5pt,
linewidth=1pt]{dBox}

\newmdenv[skipabove=7pt,
skipbelow=7pt,
backgroundcolor=darkred!15,
innerleftmargin=5pt,
innerrightmargin=5pt,
innertopmargin=5pt,
leftmargin=0cm,
rightmargin=0cm,
innerbottommargin=5pt,
linewidth=1pt]{rBox}

\newcommand{\nc}{\newcommand}
\nc{\bra}[1]{\langle#1|}
\nc{\ket}[1]{|#1\rangle}
\nc{\ketbra}[2]{|#1\rangle\!\langle#2|}
\nc{\braket}[2]{\langle#1|#2\rangle}

\nc{\proj}[1]{| #1\rangle\!\langle #1 |}
\nc{\avg}[1]{\langle#1\rangle}
\nc{\rank}{\operatorname{Rank}}
\nc{\smfrac}[2]{\mbox{$\frac{#1}{#2}$}}
\nc{\tr}{\operatorname{tr}}
\nc{\ox}{\otimes}
\nc{\dg}{\dagger}
\nc{\dn}{\downarrow}
\nc{\cA}{{\cal A}}
\nc{\cB}{{\cal B}}
\nc{\cC}{{\cal C}}
\nc{\cD}{{\cal D}}
\nc{\cE}{{\cal E}}
\nc{\cF}{{\cal F}}
\nc{\cG}{{\cal G}}
\nc{\cH}{{\cal H}}
\nc{\cI}{{\cal I}}
\nc{\cJ}{{\cal J}}
\nc{\cK}{{\cal K}}
\nc{\cL}{{\cal L}}
\nc{\cM}{{\cal M}}
\nc{\cN}{{\cal N}}
\nc{\cO}{{\cal O}}
\nc{\cP}{{\cal P}}
\nc{\cQ}{{\cal Q}}
\nc{\cR}{{\cal R}}
\nc{\cS}{{\cal S}}
\nc{\cT}{{\cal T}}
\nc{\cU}{{\cal U}}
\nc{\cV}{{\cal V}}
\nc{\cX}{{\cal X}}
\nc{\cY}{{\cal Y}}
\nc{\cZ}{{\cal Z}}
\nc{\cW}{{\cal W}}
\nc{\csupp}{{\operatorname{csupp}}}
\nc{\qsupp}{{\operatorname{qsupp}}}
\nc{\var}{{\operatorname{var}}}
\nc{\rar}{\rightarrow}
\nc{\lrar}{\longrightarrow}
\nc{\polylog}{{\operatorname{polylog}}}
\nc{\wt}{{\operatorname{wt}}}
\nc{\supp}{{\operatorname{supp}}}

\nc{\argmin}{{\operatorname{argmin}}}
\nc{\pd}[2]{\frac{\partial #1}{\partial #2}}

\def\x{\xi}

\nc{\RR}{{{\mathbb R}}}
\nc{\CC}{{{\mathbb C}}}
\nc{\FF}{{{\mathbb F}}}
\nc{\NN}{{{\mathbb N}}}
\nc{\ZZ}{{{\mathbb Z}}}
\nc{\PP}{{{\mathbb P}}}
\nc{\QQ}{{{\mathbb Q}}}
\nc{\UU}{{{\mathbb U}}}
\nc{\EE}{{{\mathbb E}}}
\nc{\id}{{\operatorname{id}}}

\nc{\CHSH}{{\operatorname{CHSH}}}

\newcommand{\Op}{\operatorname}
\nc{\rU}{\mbox{U}}

\nc{\ob}[1]{#1}
\nc{\Var}[1]{\Op{Var}[#1]}

\nc{\SEP}{{\text{\rm SEP}}}
\nc{\NS}{{\text{\rm NS}}}
\nc{\LOCC}{{\text{\rm LOCC}}}
\nc{\PPT}{{\text{\rm PPT}}}
\nc{\EXT}{{\text{\rm EXT}}}
\nc{\Sym}{{\operatorname{Sym}}}

\nc{\ERLO}{{E_{\text{r,LO}}}}
\nc{\ERLOCC}{{E_{\text{r,LOCC}}}}
\nc{\ERPPT}{{E_{\text{r,PPT}}}}
\nc{\ERLOCCinfty}{{E^{\infty}_{\text{r,LOCC}}}}
\nc{\Aram}{{\operatorname{\sf A}}}

\newcommand{\eps}{\varepsilon}

\usepackage{hyperref}
\hypersetup{colorlinks=true,citecolor=blue,linkcolor=blue,filecolor=blue,urlcolor=blue,breaklinks=true}

\makeatletter
\def\grd@save@target#1{%
  \def\grd@target{#1}}
\def\grd@save@start#1{%
  \def\grd@start{#1}}
\tikzset{
  grid with coordinates/.style={
    to path={%
      \pgfextra{%
        \edef\grd@@target{(\tikztotarget)}%
        \tikz@scan@one@point\grd@save@target\grd@@target\relax
        \edef\grd@@start{(\tikztostart)}%
        \tikz@scan@one@point\grd@save@start\grd@@start\relax
        \draw[minor help lines,magenta] (\tikztostart) grid (\tikztotarget);
        \draw[major help lines] (\tikztostart) grid (\tikztotarget);
        \grd@start
        \pgfmathsetmacro{\grd@xa}{\the\pgf@x/1cm}
        \pgfmathsetmacro{\grd@ya}{\the\pgf@y/1cm}
        \grd@target
        \pgfmathsetmacro{\grd@xb}{\the\pgf@x/1cm}
        \pgfmathsetmacro{\grd@yb}{\the\pgf@y/1cm}
        \pgfmathsetmacro{\grd@xc}{\grd@xa + \pgfkeysvalueof{/tikz/grid with coordinates/major step}}
        \pgfmathsetmacro{\grd@yc}{\grd@ya + \pgfkeysvalueof{/tikz/grid with coordinates/major step}}
        \foreach \x in {\grd@xa,\grd@xc,...,\grd@xb}
        \node[anchor=north] at (\x,\grd@ya) {\pgfmathprintnumber{\x}};
        \foreach \y in {\grd@ya,\grd@yc,...,\grd@yb}
        \node[anchor=east] at (\grd@xa,\y) {\pgfmathprintnumber{\y}};
      }
    }
  },
  minor help lines/.style={
    help lines,
    step=\pgfkeysvalueof{/tikz/grid with coordinates/minor step}
  },
  major help lines/.style={
    help lines,
    line width=\pgfkeysvalueof{/tikz/grid with coordinates/major line width},
    step=\pgfkeysvalueof{/tikz/grid with coordinates/major step}
  },
  grid with coordinates/.cd,
  minor step/.initial=.2,
  major step/.initial=1,
  major line width/.initial=2pt,
}
\makeatother

\usepackage{thmtools}
\usepackage{thm-restate}
\usepackage{etoolbox}
\makeatletter
\def\problem@s{}
\newcounter{problems@cnt}

\newcommand{\allproblems}{\problem@s}
\makeatother